\documentclass{jfm}

\usepackage{graphicx}
\usepackage{newtxtext}
\usepackage{newtxmath}
\usepackage{natbib}
\usepackage{hyperref}
\usepackage{bm,physics,booktabs,cancel,mathtools}
\hypersetup{
    colorlinks = true,
    urlcolor   = blue,
    citecolor  = black,
}

\newcommand{\RomanNumeralCaps}[1]
\linenumbers

\newcommand{\innprod}[2]{\left\langle#1,\,#2\right\rangle}
\newcommand{\ii}{\mathrm{i}\mkern1mu}

\title{Searching for Invariant Solutions to Wall-Bounded Flows using Resolvent-Based Optimisation}

\author{Thomas Burton\aff{1}
  \corresp{\email{T.P.Burton@soton.ac.uk}},
  Sean Symon\aff{1}
 \and Davide Lasagna\aff{1}}

\affiliation{\aff{1}University of Southampton, Faculty of Engineering and Physical Sciences, University Road, Southampton SO17 1BJ, UK}

\begin{document}
\maketitle

\begin{abstract}
We present a robust optimisation framework for computing invariant solutions of wall-bounded flows by recasting the Navier–Stokes equations as a variational problem as established in \mbox{Ashtari \& Schneider, JFM (2023)}. The approach minimises the residual of the governing equations over a finite time horizon, seeking periodic or equilibrium solutions. A novel contribution is made by including a Galerkin projection onto a basis of divergence-free modes that satisfy the no-slip boundary conditions. This projection not only makes the variational framework applicable to wall-bounded flows but it also yields a low-order representation of the dynamics. The basis is derived from resolvent analysis, which provides an orthonormal set. We demonstrate the method on a streamwise invariant formulation of rotating plane Couette flow, obtaining exact equilibrium and periodic solutions consistent with direct numerical simulations. The conditioning of the optimisation problem is analysed in detail, showing that convergence rates depend on the stability properties of the targeted solutions. Finally, we highlight a direct link between the conditioning of the optimisation and the structure of the resolvent operator, suggesting a unifying perspective on both the efficiency of the optimisation and the dynamical significance of resolvent modes.
\end{abstract}


\section{Introduction}\label{sec:introduction}
Turbulence is often treated as a stochastic process. Progress in its study has largely come from statistical descriptions that exploit the self-similarity across different scales, yielding a universal picture of small-scale dynamics. Such approaches, however, offer little insight into the geometry-dependent large-scale structures or the physical mechanisms that govern turbulence more generally. An alternative view of turbulence, motivated by the theory of chaos for low-dimensional dissipative systems, was proposed by Eberhard Hopf \citep{hopf1942,hopf1948}. In this view, the state of a fluid is asymptotically confined to a finite-dimensional invariant subspace embedded within the infinite-dimensional state-space of the flow. The dimension of this invariant subspace can be much smaller than the infinite-dimensional nature of the state would suggest, owing to the strong dissipative power of viscosity. As the Reynolds number increases the dimension of the invariant subspace usually increases allowing for more complex temporal evolution of the flow and requiring more degrees of freedom to accurately describe the resulting smaller scale motions observed. However, understanding this invariant subspace is quite difficult as a result of its complex fractal structure indicative of strange attractors \citep{ruelle1971,aizawa1982}.

For low-dimensional chaotic systems, the complex structure of strange attractors has been shown in \citet{cvitanovic1987} to be determined by the set of Unstable Periodic Orbits (UPOs) embedded within the attractor, providing a ``skeleton'' for the dynamics. It was shown in \citet{cvitanovic1988} and \citet{cvitanovic1995} that building these UPOs up in the form of a weighted sum can yield the ergodic properties of the dynamics on the strange attractor \citep{artuso1990a}. This method is called cycle expansion theory. Analogously, exact nonlinear solutions to the Navier-Stokes equations have been numerically shown to exist, dubbed Exact Coherent Structures (ECSs) due to their close resemblance to coherent structures observed in experiment and Direct Numerical Simulation (DNS) studies. The exact nature of these ECSs is not yet fully understood, however, there is evidence that turbulent flows repeatedly shadow ECSs for a finite amount of time before moving away along one of the unstable manifolds of the solution towards a different ECSs \citep{suri2020,krygier2021,crowley2022}. It was also demonstrated \citep{yalniz2021,page2024a} that using ECSs in a cycle expansion approach yields accurate statistics of the turbulent flow. The results do however generally rely on a tuning of the relative contribution of each ECS as opposed to deriving them exactly according cycle expansion theory \citep{wang2025}. The numerical computation of ECSs was first shown in \citet{nagata1990} and \citet{nagata1997} where equilibrium and travelling wave solutions to the plane-Couette flow were computed. Since then there has been regular work to expand the sets of solutions for various flows \citep{waleffe1998,waleffe2001,faisst2003,wedin2004,viswanath2007,pringle2007,gibson2008,waleffe2009,itano2009,wedin2009,okino2010,uhlmann2010,willis2013,nagata2021}. A general review of the relevance of invariant solutions to fluid turbulence can be found in \citet{kawahara2012}.

The numerical methods employed to find ECSs are generally divided into local and global approaches. Local methods, primarily shooting algorithms, are initialised with a flow snapshot that is evolved forward over the current period, with the mismatch between initial and final states used to update both the initial condition and the period. Examples include \citet{christiansen1997}, \citet{tomoaki2000}, \citet{kawahara2001}, \citet{sanchez2004}, \citet{veen2006}, \citet{cvitanovic2010}, and \citet{chandler2013}. Shooting algorithms are typically combined with a Newton–Krylov iteration, where the Jacobian system is solved by GMRES \citep{saad1986} to avoid explicit formation and prohibitive memory costs. However, this approach is highly sensitive to initial conditions due to the chaotic nature of the dynamics. { Robust methods to generate initial guesses include close recurrences generated from time integration \citep{viswanath2007,schneider2022,redfern2024}, project-then-search \citep{sharma2020}, or Reynolds continuation from known solutions \citep{schneider2023}, all of which become more difficult to use for longer periods and higher-dimensional problems.} To address this, multiple-shooting methods \citep{christiansen1997} replace the single trajectory with shorter segments matched simultaneously, improving conditioning and enabling parallel computation \citep{sanchez2004}. Further robustness is achieved by adding a hookstep to the Newton–GMRES iteration \citep{dennis1996,viswanath2007,viswanath2009}, which, when combined with multiple-shooting methods, represents the most effective way of computing invariant solutions in fluid flows. Nevertheless, challenges remain as the system size grows \citep{veen2019}.

Global methods instead act on a temporally extended field that already obeys the time-periodicity constraint, rather than on a set of sequential field snapshots obtained from time integration. This space-time field is then iteratively modified until its spatial variation and temporal evolution together satisfy the governing equations to within a prescribed tolerance. The Newton flow method introduced by \citet{lan2004} formulates the problem through variational dynamics where the solution residual decays exponentially (in the limit of continuous deformation) to invariant solutions, with an over-relaxation factor mitigating the erratic behaviour of standard Newton iterations. Unlike shooting methods, these global approaches avoid exponential trajectory divergence and are generally more robust. However, the Newton flow method is not naturally matrix-free, as each iteration requires forming and solving an $N\times N$ Jacobian system, with $N$ the degrees of freedom for both space and time. This makes the approach prohibitively costly for high-dimensional systems \citep{fazendeiro2010}. Practical implementations must therefore employ GMRES, as in local methods. Despite these advantages, the method still suffers from sensitivity to initial guesses owing to the limited convergence radius of Newton iterations.

Fourier–Galerkin (or harmonic-balance) formulations offer an efficient global approach for computing time-periodic invariant solutions without long integrations. By projecting the Navier–Stokes equations onto a truncated Fourier basis in time, the problem reduces to a nonlinear algebraic system for the harmonic coefficients. Recent work has demonstrated the effectiveness of this strategy for periodic and quasi-periodic flows \citep{sierraaustin2022}, its extension to adjoint-based sensitivity and stability analysis \citep{sierra2021}, and its application to nonlinear input–output analysis capturing triadic energy transfers \citep{rigas2021}. These Fourier–Galerkin approaches provide a complementary route to invariant solutions, trading time-marching for a frequency-domain nonlinear solve while naturally incorporating harmonic interactions and sensitivities.

An alternative global approach is to recast the search for invariant solutions as an optimisation problem rather than root finding. Here, the objective functional measures the total violation of the governing equations over the spatio-temporal domain. Introduced by \citet{farazmand2016} for equilibrium and travelling-wave solutions in 2D Kolmogorov flow, this method is formulated as the adjoint of the Newton descent of \citet{lan2004}. Crucially, it is naturally matrix-free and therefore well-suited to high-dimensional systems. The variational optimisation approach demonstrated markedly greater robustness to initial conditions, enabling convergence from a wider set of guesses and uncovering previously unknown solutions. More recently, \citet{schneider2022} extended the framework to periodic flows, and \citet{schneider2023} applied it to finding equilibrium in wall-bounded problems by enforcing no-slip constraints through the Influence Matrix (IM) method \citep{kleiser1980}, which maintains consistency with the  pressure-Poisson equation and avoids boundary errors arising from the lack of an explicit pressure condition.

Despite its advantages, the variational optimisation method still faces challenges. While the IM method provides a clear strategy for equilibria in wall-bounded domains, its extension to temporally periodic fields has not yet been published in the literature, leaving open the problem of a general approach for periodic solutions in such settings. Moreover, the improved robustness compared with Newton-based methods comes at the cost of losing quadratic convergence. Convergence is initially rapid but slows considerably near the minimum, largely due to the use of gradient descent, which is notoriously inefficient in this regime \citep{nocedal2006}. Algorithms incorporating curvature information could improve performance. \citet{farazmand2016} addressed this by adopting a hybrid strategy, initially using optimisation to approach the solution after which Newton–GMRES–hookstep is employed to accelerate convergence. More recently, \citet{schneider2023} improved efficiency of variational optimisation by employing an extrapolation technique with Dynamic Mode Decomposition (DMD). Nevertheless, the method remains hampered by a significantly slower convergence rate than Newton-based methods, particularly as problem dimension increases. Hence, it is desirable to develop modifications that improve convergence without switching methods, while retaining compatibility with complex no-slip geometries.

In this work, the variational optimisation framework is extended with a Galerkin projection, providing a general methodology for wall-bounded periodic flows with no-slip boundaries. The form of this problem closely resembles the general harmonic balance method described in \citet{sierraaustin2022}. This projection enforces incompressibility and no-slip constraints while decoupling pressure and velocity. The basis is constructed with resolvent modes, which form a divergence-free orthonormal set satisfying the boundary conditions. Our approach is similar to \citet{li2025}, who used a projection-based optimisation for lid-driven cavity flow  using modes derived from Spectral Proper Orthogonal Decomposition (SPOD). In this work, however, resolvent modes replace these SPOD modes, removing the need for fully resolved flow data and enabling the construction of exact solutions rather than reduced-order models. Resolvent analysis, a widely used tool for stability, control, and modelling \citep{gayme2010,garnaud2013,gomez2016,beneddine2017,symon2018,gayme2019,jin2022}, has been shown to provide dynamically significant bases for unstable solutions \citep{sharma2016,rosenberg2019b} and to enable new invariant solutions via “projection-then-search” \citep{sharma2020}. It has also been applied to reduced-order modelling, e.g. Taylor vortex flow in \citet{rosenberg2019a} and \citet{barthel2021}. Our work builds on these ideas by employing resolvent modes within a variational framework, where the Galerkin projection both enforces boundary constraints and effectively pre-conditions the optimisation. In this setting, modal truncation is used not merely to approximate turbulence, as in \citet{barthel2021}, but to enable computation of exact invariant solutions while simultaneously improving the conditioning of the optimisation and thereby accelerating convergence. To further address the slow convergence inherent to variational optimisation, we utilise two strategies. First, we replace the gradient-descent schemes of \citet{farazmand2016,schneider2022,schneider2023} with quasi-Newton algorithms such as L-BFGS and conjugate gradient, which incorporate curvature information. We demonstrate a substantially faster convergence rate near the minimum compared to gradient descent.  Second, we exploit the typically low-dimensional representation of ECSs using resolvent modes \citep{sharma2016} to truncate the basis, restricting the optimisation to a smaller space. It is then demonstrated that the convergence rate of the optimisation is improved via this truncation, a result that is motivated by bounding the conditioning of the optimisation problem to the singular values of the resolvent operator. This dual use of the projection framework thus addresses both the difficulty of enforcing no-slip constraints in wall-bounded flows and the slow convergence that has limited the practical utility of variational optimisation.

The remainder of the paper is organised as follows. Section~\ref{sec:methodology} introduces the variational optimisation methodology and the Galerkin projection used to enforce incompressibility and no-slip boundary conditions. Section~\ref{sec:resolvent} presents the derivation of the resolvent modes forming the projection basis. Section~\ref{sec:rpcf} introduces the 2D3C formulation of Rotating Plane Couette Flow (RPCF) and features a basic analysis of the behaviour of the flow at various Reynolds number regimes. Section~\ref{sec:solutions} then demonstrates the projected variational optimisation methodology using resolvent analysis, featuring equilibrium and periodic solutions. The underlying mechanics governing the conditioning of the optimisation is then discussed in section~\ref{sec:conditioning}, providing a novel link between resolvent analysis and the convergence rate related to the formulation discussed in \citet{mons2021}. Finally, section~\ref{sec:conclusions} concludes with a discussion of the main results of this work and the possible avenues for future work.

\section{Methodology}\label{sec:methodology}
In section~\ref{sec:variational-solver}, the variational optimisation methodology for wall-bounded flows is described.  The methodology is specialised for planar Couette flow. In section~\ref{sec:galerkin-projection} the Galerkin projection is introduced as a way to solve to issues that arise due to the presence of no-slip boundary conditions. A more general derivation and discussion of the implications can be found in \citet{burton2025-thesis}.

\subsection{Variational Optimisation}\label{sec:variational-solver}

\begin{figure}
    \centering
    \includegraphics[width=0.45\textwidth]{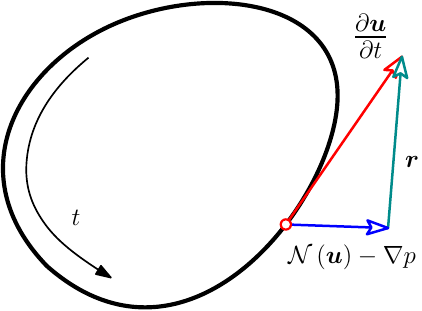}
    \caption{Schematic of an arbitrary loop in state-space that does not satisfy the governing equations as its tangent vector $\partial\bm{u}/\partial t$ is not aligned with the evolution operator $\mathcal{N}=-\left(\bm{u}\cdot\bm{\nabla}\right)\bm{u}+\frac{1}{\mathit{Re}}\bm{\Delta}\bm{u}$.}
    \label{fig:var-optimisation}
\end{figure}

 Consider a domain bounded by two parallel plates, denoted as $\Omega$, with a single inhomogeneous wall-normal direction ($y$) normalised with respect to the half channel height denoted as $h$, and the streamwise and spanwise directions ($x$ and $z$, respectively) which possess periodic boundary conditions. The unit vectors in each direction are denoted as $\bm{e}_x$, $\bm{e}_y$, and $\bm{e}_z$. The periodic boundary conditions model the effect of an infinite domain in these directions which result in statistically homogeneous turbulence. The domain can then be defined as $\Omega=[0,\,L_x]\times[-1,\,1]\times[0,\,L_z]$ with $L_x$ and $L_z$ denoting the streamwise and spanwise length of the domain, respectively, normalised with respect to the channel height. The velocity $\bm{u}$ is a $3$-dimensional vector field defined over the $3$-dimensional space $\Omega$, and the pressure field, denoted with $p$, is a scalar field over the same domain. The flow in this domain is governed by (non-dimensional) incompressible Navier-Stokes equations and continuity equation
\begin{subequations}
    \begin{align}
        \pdv{\bm{u}}{t}&=-\grad{p}-\left(\bm{u}\cdot\bm{\nabla}\right)\bm{u}+\frac{1}{\mathit{Re}}\bm{\Delta}\bm{u}, \label{eq:dynamical-system} \\
        \div{\bm{u}}&=0,
    \end{align}
    \label{eq:general-ns}%
\end{subequations}
where $\mathit{Re}=U_wh/\nu$ where $U_w$ is the speed of the walls, and $\nu$ is the kinematic viscosity of the fluid. The Laplace operator of functions in this domain is defined as $\bm{\Delta}=\partial^2/\partial x^2+\partial^2/\partial y^2+\partial^2/\partial z^2$. The no-slip boundary conditions are imposed at the walls
\begin{equation}
    \left.\bm{u}\right|_{y=\pm1}=\pm\bm{e}_x.
    \label{eq:general-no-slip}
\end{equation}
The velocity field $\bm{u}$ is also defined over a finite time horizon fixed by the orbit period $T$, thus $t\in\left[0,\,T\right)$. The combination of the spatial and temporal domains is defined as $\Omega_t=\Omega\cross\left[0,\,T\right)$. The  inner product on this space is defined as
\begin{equation}
    \innprod{\bm{u}}{\bm{v}}_{\Omega_t}=\int_0^T\int_0^{L_z}\int_{-1}^1\int_0^{L_x}\bm{u}\vdot\bm{v}\dd{x}\dd{y}\dd{z}\dd{t}.
    \label{eq:inner-product}
\end{equation}
 The operator $\vdot$ denotes the Euclidean dot product between two finite vectors. For real valued vectors this given as $\bm{u}\vdot\bm{v}=\bm{u}^\top\bm{v}$ where $(\cdot)^\top$ denotes the transpose of the a vector, and for complex valued vectors it is $\bm{u}\vdot\bm{v}=\bm{u}^\dagger\bm{v}$ where $(\cdot)^\dagger$ denotes the conjugate transpose of the vector. This  inner product induces an associated norm $\norm{\bm{u}}_{\Omega_t}=\sqrt{\innprod{\bm{u}}{\bm{u}}_{\Omega_t}}$. Thus, any velocity  field that has a finite norm under this definition is an element of the underlying Hilbert space, denoted with $\chi$, and has finite kinetic energy. We now define the space of all periodic incompressible flow fields that obey the boundary and periodicity conditions
\begin{equation}
    \mathcal{P}_T=\left\{\bm{u}\in\chi\,:\,\div{\bm{u}}=0,\,\left.\bm{u}\right|_{t=0}=\left.\bm{u}\right|_{t=T},\,\left.\bm{u}\right|_{y=\pm1}=\pm\bm{e}_x,\,\bm{u}\text{obeys periodic BCs}\right\}.
    \label{eq:opt-space}
\end{equation}
The elements of $\mathcal{P}_T$ are state-space loops that do not necessarily obey \eqref{eq:dynamical-system}. In general $T$ is not known a priori and should be included as part of the search for a solution. The goal is to derive a method that converges to elements of the subset of $\mathcal{P}_T$ that obey \eqref{eq:dynamical-system}. This can be done by introducing a mapping $\mathcal{R}\colon\mathcal{P}_T\to\mathbb{R}_{\geq0}$ such that $\mathcal{R}\left[\bm{u}\right]=0$ if and only if $\bm{u}\in\mathcal{P}_T$ and $\bm{u}$ satisfies \eqref{eq:dynamical-system}; otherwise, $\mathcal{R}>0$. This is equivalent to defining an objective function whose global minima ($\mathcal{R}=0$) correspond to periodic solutions of \eqref{eq:dynamical-system} that are divergence-free and obey the no-slip and periodicity boundary conditions, called here the global residual. To obtain the global residual, first a local residual that quantifies the violation of the Navier-Stokes equations at every point in $\Omega_t$ is defined as follows
\begin{equation}
    \bm{r}=\pdv{\bm{u}}{t}+\grad{p}+\left(\bm{u}\cdot\grad{\bm{u}}\right)-\frac{1}{\mathit{Re}}\bm{\Delta}\bm{u}.
    \label{eq:residual}
\end{equation}
Figure~\ref{fig:var-optimisation} shows a representation of the local residual defined on all points of a state-space loop representing a periodic flow. The local residual spans the distance between the rate of change of a state-space loop and the forcing imposed by the Navier-Stokes equations. The task of finding periodic solutions to \eqref{eq:dynamical-system} in the space $\mathcal{P}_T$ can now be stated in terms of the following optimisation problem
\begin{equation}
    \min_{\bm{u}\in\mathcal{P}_T,\,T}\quad\mathcal{R}\left[\bm{u}\right]=\frac{1}{2}\norm{\bm{r}}^2_{\Omega_t}.
    \label{eq:opt-prob}
\end{equation}
Geometrically, minimising $\mathcal{R}$ is equivalent to continuously deforming a state-space loop as depicted in figure~\ref{fig:var-optimisation} such that the vectors $\pdv*{\bm{u}}{t}$ and $\grad{p}+\left(\bm{u}\cdot\grad{\bm{u}}\right)-\frac{1}{\mathit{Re}}\bm{\Delta}\bm{u}$ align as closely as possible, while being constrained to the linear subspace that define the boundary conditions and incompressibility conditions. This process terminates when the rate of change of the state vector and the right-hand side of \eqref{eq:dynamical-system} are as closely aligned as possible.

To be able to solve this optimisation problem in practice, the gradient of $\mathcal{R}$ with respect to the field $\bm{u}$ is required, referred to as the functional derivative and denoted as $\fdv*{\mathcal{R}}{\bm{u}}$. This expression can be obtained by adding a perturbation to the velocity field denoted as $\bm{v}$ such that $\bm{u}\to\bm{u}+\epsilon\bm{v}$ where $\epsilon\in\mathbb{R}$. Substituting this into the definitions of the global residual relates the change in the residual, called the first variation of $\mathcal{R}$ and denoted as $\delta\mathcal{R}$, to the desired residual gradient by
\begin{equation}
    \delta\mathcal{R}=\innprod{\fdv{\mathcal{R}}{\bm{u}}}{\epsilon\bm{v}}_{\Omega_t}.
\end{equation}
Linearising the local residual in \eqref{eq:residual} and taking the adjoint of the resulting linearised operators gives the closed-form expression for the residual gradient
\begin{equation}
    \fdv{\mathcal{R}}{\bm{u}}=-\pdv{\bm{r}}{t}-\left(\bm{u}\vdot\grad\right)\bm{r}+\left(\grad{\bm{u}}\right)^\top\bm{r}-\frac{1}{\mathit{Re}}\bm{\Delta}\bm{r}-\grad{q},
    \label{eq:residual-grad}
\end{equation}
with the additional constraints on the local residual
\begin{subequations}
    \begin{align}
        \div{\bm{r}}&=0, \\
        \left.\bm{r}\right|_{y=\pm1}&=0,
    \end{align}
    \label{eq:adjoint-constraints}%
\end{subequations}
as well as $\bm{r}$ obeying the periodic boundary conditions. The details of the derivation for \eqref{eq:residual-grad} and the source of the extra constraints in \eqref{eq:adjoint-constraints} are given in appendix~\ref{app:residual-grad-wrt-velocity}.

To implement a gradient-based optimisation of \eqref{eq:opt-prob} it is also necessary to have a gradient of $\mathcal{R}$ with respect to $T$. In this work, however, the fundamental frequency, defined as $\omega=2\pi/T$, is preferred. The fundamental frequency represents the lowest frequency oscillation that is permissible in the finite time window defined by the period $T$. The gradient of $\mathcal{R}$ with respect to $\omega$ is given by
\begin{equation}
    \pdv{\mathcal{R}}{\omega}=\frac{1}{\omega}\innprod{\pdv{\bm{u}}{t}}{\bm{r}}_{\Omega_t},
    \label{eq:frequency-gradient}
\end{equation}
of which the detailed derivation is given in appendix~\ref{app:residual-grad-wrt-frequency}.

The derivations described in \citet{farazmand2016}, \citet{schneider2022}, and \citet{schneider2023} formulate the problem as a new set of dynamics which leads to taking the adjoint of the linearised dynamics. This is the primary reason that the method is typically known as the adjoint solver method rather than variational optimisation. The introduction of the new variable $q$ in \eqref{eq:residual-grad} is required to enforce the constraint that $\div{\fdv*{\mathcal{R}}{\bm{u}}}=0$ which ensures that as the optimisation progresses the flow remains incompressible. Since $q$ performs an identical role to the pressure $p$ in the primitive dynamics, we call it the adjoint pressure in line with its role in \citet{schneider2023}. The expression \eqref{eq:residual-grad} is the same as that derived in \citet{schneider2023}, with the adjoint pressure $q$ arising as a consequence of including the continuity equation as part of the constructed adjoint dynamics instead of an explicit constraint as it is treated here. Some of the general features of the derived variational/adjoint dynamics and its treatment in \citet{schneider2023} are discussed in appendix~\ref{app:dae}.

\subsection{Galerkin Projection}\label{sec:galerkin-projection}
The difficulty in solving the optimisation problem in \eqref{eq:opt-prob} comes from not having a simple way to compute the gradient in \eqref{eq:residual-grad} while enforcing the constraints on the local residual given in \eqref{eq:adjoint-constraints}. This stems from the lack of physical boundary conditions for the pressures $p$ and $q$. To provide a na\"ive example, if one initialises an optimisation with a candidate field $\bm{u}\in\mathcal{P}_T$ with period $T$ that does not satisfy \eqref{eq:dynamical-system}, then to compute $\bm{r}$ in \eqref{eq:residual} it is required to first find the pressure field $p$. This would typically be done by solving a  pressure-Poisson equation with Neumann boundary conditions which ensures that $\div{\bm{r}}=0$. However, it does not guarantee $\left.\bm{r}\right|_{y=\pm1}=0$ since the Dirichlet boundary conditions are not satisfied by default, due to $\bm{u}$ not representing an actual solution to the Navier-Stokes equations. The result of this procedure is non-zero residual values at the wall, which invalidates the gradient $\fdv*{\mathcal{R}}{\bm{u}}$ as a guaranteed descent direction for $\mathcal{R}$, due to the constraints in \eqref{eq:adjoint-constraints} not being properly satisfied. In addition, the same problem exists for computing the adjoint pressure $q$ leading to a similar problem where it is very difficult to simultaneously enforce $\div{\bm{u}}=0$ and  $\left.\bm{u}\right|_{y=\pm1}=\pm\bm{e}_x$. This leads to a velocity field at the next iteration not obeying the no-slip boundary conditions. Put another way, if the system of equations that make up the variational dynamics is solved using the  pressure-Poisson equations for $p$ and $q$ with standard Neumann boundary conditions, then the no-slip boundary conditions cannot be explicitly enforced and $\bm{u},\,\bm{r}\notin\mathcal{P}_T$.

\citet{schneider2023} proposed a methodology specific to equilibrium solutions that avoids this issue by solving for both the update to the velocity field, $\bm{u}$, and the pressure field, $p$, simultaneously in a coupled fashion using the Influence Matrix (IM) method, which ensures that both $\div{\bm{r}}=0$ and $\left.\bm{r}\right|_{y=\pm1}=0$ are satisfied. The same treatment is provided for $\bm{r}$ and $q$ in a staggered approach, providing a physically consistent evolution of both $\bm{u}$ and $\bm{r}$. However, this method was developed to solve for equilibria ($\pdv*{t}=0$), and it is not obvious how this would work for temporally varying fields because the method used to obtain compatible velocity and pressure fields is designed as an update method for some time-stepping scheme,  while we employ a Fourier basis in time. 

With the difficulties in evolving the gradient-based optimisation while satisfying all the constraints now clear, we propose to use a Galerkin projection onto an orthogonal basis that satisfies the boundary conditions and the incompressibility constraints. Some setup is required to arrive at the final procedure that yields a valid computation of the residuals and gradient that obey the constraints. First, the velocity field is expanded into a sum over a basis as follows
\begin{equation}
    \bm{u}\left(x,\,y,\,z,\,t\right)=\bm{u}_{\text{BC}}(y)+\sum_{m=1}^\infty\sum_{\bm{k}\in\mathbb{Z}^3}a_{\bm{k}m}\bm{\psi}_{\bm{k}m}\left(y\right)e^{{\ii}\bm{k}\cdot\bm{\xi}},
    \label{eq:velocity-expansion}
\end{equation}
where  $\bm{k}=\left(k_x,\,k_z,\,k_t\right)$ is the integer wavenumber vector, and  $\bm{\xi}=\left(\alpha x,\,\beta z,\,\omega t\right)$ is the scaled direction vector. The coefficients $\alpha=2\pi/L_x$, $\beta=2\pi/L_z$, and $\omega=2\pi/T$ are determined by the domain size and period, and each represent the smallest frequency in each of their respective direction that can be accommodated in the finite space.  To account for the inhomogeneous boundary conditions on $\bm{u}$ in \eqref{eq:general-no-slip}, the steady base flow $\bm{u}_{\text{BC}}\left(y\right)$ is introduced into \eqref{eq:velocity-expansion}. In this work, the laminar state will be used for the base flow. The modes $\bm{\psi}_{\bm{k}m}=\bm{\psi}_{\bm{k}m}\left(y\right)$ are divergence-free and obey the no-slip boundary conditions
\begin{align}
    \grad_{\bm{k}}\vdot\bm{\psi}_{\bm{k}m}=0, \label{eq:mode-incompressible}\\
    \left.\bm{\psi}_{\bm{k}m}\right|_{y=\pm1}=0, \label{eq:mode-ns}
\end{align}
in addition to being orthonormal
\begin{equation}
    \int_{-1}^1\bm{\psi}_{\bm{k}m}\vdot\bm{\psi}_{\bm{k}n}\dd{y}=\delta_{mn},
    \label{eq:mode-orthonormal}
\end{equation}
 where $\delta_{mn}$ is the Kronecker delta. The operator  $\grad_{\bm{k}}\vdot\bm{u}_{\bm{k}}(y)=i\alpha k_x\bm{u}_{\bm{k}}+\partial\bm{u}_{\bm{k}}/\partial y+i\beta k_z\bm{u}_{\bm{k}}$ is the divergence operator in spectral/Fourier space. The sum over the set of modes $\bm{\psi}_{\bm{k}m}$ is taken over a countably infinite set of modes for every frequency $\bm{k}$ to form a complete basis for the optimisation space $\mathcal{P}_T$. This includes the mean mode $\bm{k}=\bm{0}$ as this is part of the solution that needs to determined during the optimisation. In practice the sums in \eqref{eq:velocity-expansion} are truncated to be finite. The key observation is that for any combination of coefficients $a_{\bm{k}m}\in\mathbb{C}$ the velocity field is incompressible and obeys the no-slip boundary conditions, restricting the velocity field to exist only within a finite-dimensional sub-space of the complete function space.

\begin{figure}
    \centering
    \includegraphics[width=0.5\linewidth]{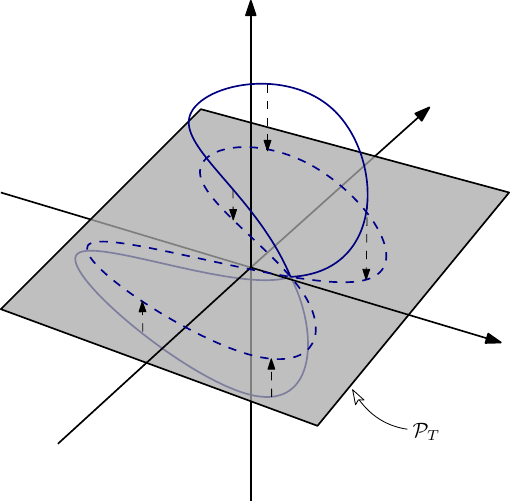}
    \caption{Schematic for a Galerkin projection of state-space loop representing a velocity field onto the linear subspace in \eqref{eq:opt-space}.}
    \label{fig:projection-diagram}
\end{figure}

The orthogonality of the modes allows for the following identity for the coefficients $a_{\bm{k}m}$ to be derived
\begin{equation}
    a_{\bm{k}m}=\frac{1}{2TL_zL_x}
    \label{eq:velocity-projection}
\end{equation}
This projection is least-squares, in the sense that the set of coefficients $a_{\bm{k}m}$ produces a trajectory that is the closest possible trajectory to $\bm{u}$ restricted to be within the linear subspace defined by $\bm{\psi}_{\bm{k}m}$. A low-dimensional schematic for such a projection is depicted in figure~\ref{fig:projection-diagram}.

Next, a similar expansion can be performed for the local residual
\begin{equation}
    \bm{r}=\sum_{m=1}^\infty\sum_{\bm{k}\in\mathbb{Z}^3}s_{\bm{k}m}\bm{\psi}_{\bm{k}m}\left(y\right)e^{\ii\bm{k}\cdot\bm{\xi}},
    \label{eq:residual-expansion}
\end{equation}
Similar to \eqref{eq:velocity-expansion}, any combination of coefficients $s_{\bm{k}m}\in\mathbb{C}$ constructs a local residual field that obeys the constraints \eqref{eq:adjoint-constraints}, making it the crucial step in this methodology. The coefficients $s_{\bm{k}m}$ can be computed in a similar way as to velocity
\begin{equation}
    s_{\bm{k}m}=\frac{1}{2TL_zL_x}\innprod{\bm{r}}{e^{-\ii\bm{k}\vdot\bm{\xi}}\bm{\psi}_{\bm{k}m}}_{\Omega_t}.
    \label{eq:residual-projection}
\end{equation}
If the expansions \eqref{eq:velocity-expansion} and \eqref{eq:residual-expansion} are substituted into \eqref{eq:residual-grad} and \eqref{eq:residual}, and the results are projected using the same operation as in \eqref{eq:velocity-projection} we get the following expressions for the gradient of $\mathcal{R}$ with respect to the coefficients $a_{\bm{k}m}$
\begin{equation}
    \pdv{\mathcal{R}}{a_{\bm{k}m}}=-ik_t\omega s_{\bm{k}m}-\frac{1}{2TL_zL_x}\innprod{\left(\bm{u}\cdot\bm{\nabla}\right)\bm{r}-\left(\grad{\bm{u}}^\top\right)\bm{r}+\frac{1}{\mathit{Re}}\bm{\Delta}\bm{r}}{e^{-\ii\bm{k}\vdot\bm{\xi}}\bm{\psi}_{\bm{k}m}}_{\Omega_t},
    \label{eq:projected-gradient}
\end{equation}
and for the local residual coefficients $s_{\bm{k}m}$
\begin{equation}
    s_{\bm{k}m}=ik_t\omega a_{\bm{k}m}+\frac{1}{2TL_zL_x}\innprod{\left(\bm{u}\cdot\bm{\nabla}\right)\bm{u}-\frac{1}{\mathit{Re}}\bm{\Delta}\bm{u}}{e^{-\ii\bm{k}\vdot\bm{\xi}}\bm{\psi}_{\bm{k}m}}_{\Omega_t}.
    \label{eq:projected-residual}
\end{equation}

\begin{figure}
    \centering
    \includegraphics{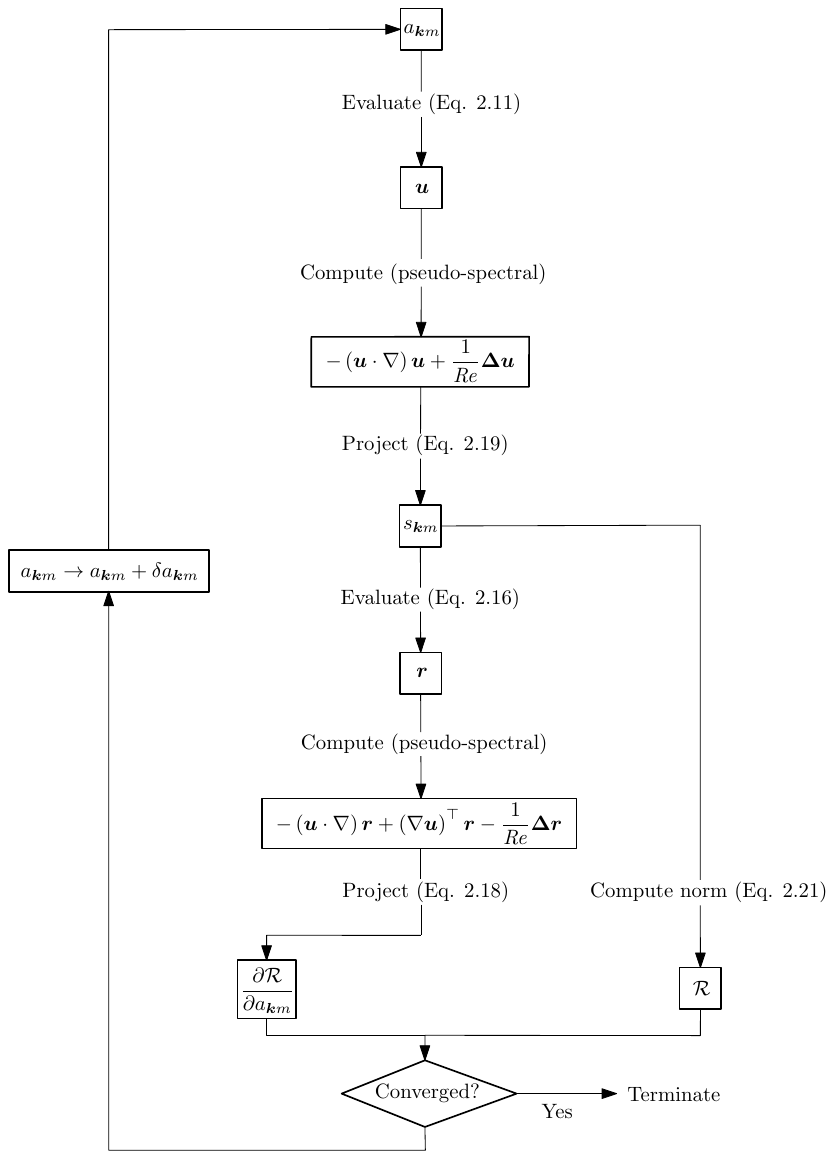}
    \caption{Flow diagram of the computations performed at each iteration of the optimisation.}
    \label{fig:optimisation-loop}
\end{figure}

The key observation to be made on the expressions \eqref{eq:projected-gradient} and \eqref{eq:projected-residual} is that the pressure gradient terms are not present. This is a result of the fact the basis $\bm{\psi}_{\bm{k}m}$ is divergence-free, leading to the following result
\begin{equation}
    \innprod{\grad{p}}{e^{-\ii\bm{k}\vdot\bm{\xi}}\bm{\psi}_{\bm{k}m}}_{\Omega_t}=\innprod{p}{-\div{\left(e^{-\ii\bm{k}\vdot\bm{\xi}}\bm{\psi}_{\bm{k}m}\right)}}_{\Omega_t}=\innprod{p}{-e^{-\ii\bm{k}\vdot\bm{\xi}}\grad_{\bm{k}}\vdot\bm{\psi}_{\bm{k}m}}_{\Omega_t}=0,
\end{equation}
and similarly for the adjoint pressure gradient term $\grad{q}$.

Figure~\ref{fig:optimisation-loop} is a flow diagram for the computations over a single iteration of a given optimisation loop, to compute both $\mathcal{R}$ and $\pdv*{\mathcal{R}}{a_{\bm{k}m}}$. The computation begins with the coefficients $a_{\bm{k}m}$ that represent a given periodic and incompressible velocity field that obeys the periodicity and no-slip boundary conditions, relative to a given set of modes $\bm{\psi}_{\bm{k}m}$ and  laminar base flow $\bm{u}_{\text{BC}}$. The residual coefficients $s_{\bm{k}m}$ are then obtained by computing all the terms in \eqref{eq:residual} excluding the pressure gradient $\grad{p}$, the result of which is then projected onto the basis as in \eqref{eq:residual-projection}. The global residual can be obtained directly from $s_{\bm{k}m}$  by substituting \eqref{eq:residual-projection} into \eqref{eq:residual} and evaluating the resulting integral, leading to the following identity
\begin{equation}
    \mathcal{R}=\frac{1}{2}\sum_{m=1}^\infty\sum_{\bm{k}\in\mathbb{Z}^3}|s_{\bm{k}m}|^2,
\end{equation}
which is simply an application of Parseval's identity. To obtain the residual gradient, the velocity and residual coefficients are expanded back into physical space to facilitate the computation of all the terms on the right-hand side of \eqref{eq:residual-grad} excluding the adjoint pressure gradient $\grad{q}$. Finally the result is projected using \eqref{eq:projected-gradient} to obtain $\pdv*{\mathcal{R}}{a_{\bm{k}m}}$. With both $\mathcal{R}$ and its gradient, the chosen optimisation algorithm can be used  to obtain an update the coefficients $a_{\bm{k}m}$, denoted as $\delta a_{\bm{k}m}$ in Figure~\ref{fig:optimisation-loop}.s Convergence being determined by the size of $\mathcal{R}$. The process summarised here allows the direct computation of the required residual and gradient without resorting to solving a  pressure-Poisson equation, avoiding the issues related to the boundary conditions. The frequency gradient $\pdv*{\mathcal{R}}{\omega}$ does not require any extra special treatment as once the appropriate local residual $\bm{r}$ has been determined then \eqref{eq:frequency-gradient} can be computed directly.


 The wall-normal derivative terms in \eqref{eq:projected-gradient} and \eqref{eq:projected-residual} can only be expressed in terms of the modes $\bm{\psi}_{\bm{k}m}$. Additionally, the computation of the nonlinear terms in projected space using $a_{\bm{k}m}$ and $s_{\bm{k}m}$ would require the evaluation of several convolution sums \citep{barthel2021} that scale poorly with the degrees of freedom of the discretised system. To avoid this the pseudo-spectral method is used. The velocity and the residual are expanded back into their full physical space forms where the nonlinear terms are computed after which they are then projected onto the modes, replacing the heavy convolution sums with pointwise multiplication in physical space.


\section{Resolvent Analysis}\label{sec:resolvent}
To implement the projected variational optimisation methodology, a method to generate the modes $\bm{\psi}_{\bm{k}m}$ with the desired properties \eqref{eq:mode-incompressible}-\eqref{eq:mode-orthonormal} is required. This choice is non-trivial with many possible solutions. For example, SPOD could be used to generate a set of modes from large data sets \citep{towne2018}, or global stability analysis can derive modes directly from the Navier-Stokes equations. Here resolvent analysis is used to generate the desired modes which form a complete basis as in \eqref{eq:velocity-expansion} and \eqref{eq:residual-expansion}. For completeness we provide an overview of resolvent analysis following closely the formulation provided in \citet{sharma2019}.

First, decompose the velocity field into a steady base component and the fluctuation around this base flow as follows
\begin{equation}
    \bm{u}\left(x,\,y,\,z,\,t\right)=\bm{u}_b\left(y\right)+\bm{u}^\prime\left(x,\,y,\,z,\,t\right),
    \label{eq:mean-fluc-decomp}
\end{equation}
where the base flow here is the flow about which the equations will be linearised, serving a distinct role from the base flow $\bm{u}_{\text{BC}}(y)$ used in \eqref{eq:velocity-expansion}. The standard choice in resolvent analysis is to take $\bm{u}_b$ to be the turbulent mean. 
The choice taken in this work is  is to use the laminar profile, primarily pragmatic reasons and is discussed further in sections~\ref{sec:solutions} and \ref{sec:conditioning}.

 At this point it is useful to define the Leray projector, denoted as $\mathbb{P}$ \citep{temam1984}. This operator takes an instantaneous velocity field and orthogonally projects onto the divergence-free subspace, and can be derived from Helmholtz decomposition of the velocity field.  It is used here as a notional short-hand which removes the pressure gradient term rendering the primitive formulation of the Navier-Stokes equation easier to manipulate. Applying the Leray projector to \eqref{eq:dynamical-system} and then substituting in the decomposition \eqref{eq:mean-fluc-decomp} provides an evolution equation for the fluctuations
\begin{equation}
    \pdv{\bm{u}^\prime}{t}=\mathbb{P}\left(\mathcal{N}\left(\bm{u}_b\right)+\mathcal{L}_{\bm{u}_b}\bm{u}^\prime+\bm{f}\left(\bm{u}^\prime\right)\right).
    \label{eq:ns-fluc}
\end{equation}
The operator $\mathcal{N}\left(\bm{u}\right)$ is the Navier-Stokes operator, given by the right-hand side of \eqref{eq:dynamical-system}. The operator $\mathcal{L}_{\bm{u}_b}$ is the linearised Navier-Stokes operator evaluated at the base flow $\bm{u}_b$. Finally, $\bm{f}\left(\bm{u}^\prime\right)=-\left(\bm{u}^\prime\vdot\grad\right)\bm{u}^\prime$ is the nonlinear term for the fluctuations $\bm{u}^\prime$. Physically $\bm{f}$ represents the nonlinear feedback that transports energy between scales and sustains any unsteady motion in the flow. The addition of the Leray projector allows us to ignore the pressure gradient terms in the operators $\mathcal{N}$ and $\mathcal{L}_{\bm{u}_b}$, as well as not needing to explicitly represent the continuity equation in \eqref{eq:ns-fluc}.

Next we define the Fourier expansion of the velocity fluctuation $\bm{u}^\prime$ in the homogeneous spatial and time directions as follows, following closely the formulation used in section~\ref{sec:galerkin-projection},
\begin{equation}
    \bm{u}^\prime\left(x,\,y,\,z,\,t\right)=\sum_{\bm{k}\in\mathbb{Z}^3}\bm{u}^\prime_{\bm{k}}\left(y\right)e^{\ii}\bm{k}\vdot\bm{\xi},
\end{equation}
where $\bm{k}$ and $\bm{\xi}$ have the same definition as in section~\ref{sec:galerkin-projection}. The corresponding inverse operation to compute the Fourier coefficients is given by
\begin{equation}
    \bm{u}^\prime_{\bm{k}}=\frac{1}{8\pi^3}\int_0^{2\pi}\int_0^{2\pi}\int_0^{2\pi}\bm{u}^\prime e^{\ii\bm{k}\vdot\bm{\xi}}\dd{\bm{\xi}}.
\end{equation}
Expanding \eqref{eq:ns-fluc} into its Fourier coefficients the following is obtained for the fluctuations
\begin{equation}
    ik_t\omega\bm{u}^\prime_{\bm{k}}=\mathbb{P}\left(\mathcal{L}_{\bm{k},\bm{u}_b}\bm{u}^\prime_{\bm{k}}+\bm{f}_{\bm{k}}\right),\quad\bm{k}\in\mathbb{Z}^3\setminus\left\{\bm{0}\right\}.
\end{equation}
Rearranging, one obtains the following linear relationship between the fluctuations and the nonlinear interactions resulting from the convective term.
\begin{equation}
    \bm{u}^\prime_{\bm{k}}=\bm{R}_{\bm{k}}\bm{f}_{\bm{k}},\quad\bm{k}\in\mathbb{Z}^3\setminus\left\{\bm{0}\right\}.
    \label{eq:resolvent}
\end{equation}
The operator $\bm{R}_{\bm{k}}=\left(ik_t\omega\bm{I}-\mathbb{P}\mathcal{L}_{\bm{k},\bm{u}_b}\right)^{-1}\mathbb{P}$ is the resolvent.
The exact form of $\bm{R}_{\bm{k}}$ is given in appendix~\ref{app:resolvent}. The next step is to decompose the operator using a Schmidt decomposition, or equivalently a Singular Value Decomposition (SVD) when discretised, as follows
\begin{equation}
    \bm{R}_{\bm{k}}\left(\cdot\right)=\sum_{m=1}^\infty\sigma_{\bm{k}m}\bm{\psi}_{\bm{k}m}\innprod{\bm{\phi}_{\bm{k}m}}{\cdot}.
    \label{eq:svd}
\end{equation}
This decomposition provides two sets of orthonormal modes, i.e.
\begin{align}
    \int_{-1}^1\bm{\psi}_{\bm{k}m}\vdot\bm{\psi}_{\bm{k}n}\dd y=\delta_{mn}, \\
    \int_{-1}^1\bm{\phi}_{\bm{k}m}\vdot\bm{\phi}_{\bm{k}n}\dd y=\delta_{mn},
\end{align}
ranked in order of the associated singular values $\sigma_m\geq\sigma_{m+1}\geq0$, for all $i\in\mathbb{N}$. The modes $\bm{\psi}_{\bm{k}m}$ and $\bm{\phi}_{\bm{k}m}$ are the left and right singular modes and form a complete basis for the range (response) and domain (input) of the resolvent for every $\bm{k}\in\mathbb{Z}^3\setminus\left\{0\right\}$, respectively. Since the range of the resolvent is the space of divergence-free fluctuation velocity fields that obey the desired boundary condition, the left singular modes, $\bm{\psi}_{\bm{k}m}$, also called the response modes, have these desired properties. Thus, the response modes $\bm{\psi}_{\bm{k}m}$ can be used as the basis for the Galerkin projection introduced in section~\ref{sec:galerkin-projection}.

To actually find exact solutions as discussed in section~\ref{sec:methodology}, a basis for the mean $\bm{k}=0$ is also required. The resolvent in \eqref{eq:resolvent} is only technically defined using the fluctuation equation at all non-zero frequencies, $\bm{k}\neq0$. However,  the authors found that the resolvent operator can regardless be evaluated at $\bm{k}=0$ which provided the required modes for the mean, completing the basis required to find exact solution using the expansions \eqref{eq:velocity-expansion} and \eqref{eq:residual-expansion}.  In other words, despite the technically undefined interpretation the authors found that taking the SVD of the resolvent at $\bm{k}=0$ yielded a valid orthonormal basis as desired.
The exact physical interpretation of the basis used for the Galerkin projection is less important  than its ability to provide a valid orthonormal basis in the space of divergence-free and no-slip fields.  The literature discussing the evaluation of the resolvent at the mean frequencies and wavenumbers is extremely sparse, since it is not typically required as the mean is used as an input. The exception is \citet{mons2021} which uses resolvent of the RANS for the purpose of data assimilation. However, this does not directly apply to the formulation in this work as the the resolvent in \eqref{eq:resolvent} is defined using the fluctuation equation not the RANS equation.

The rate at which the singular values $\sigma_m$ decay determines how accurately the decomposition \eqref{eq:svd} can be represented with only a finite number of the modes. It has been observed for fluid flows that the singular values decay rapidly \citep{mckeon2010}, meaning a low-rank approximation to the resolvent operator can be constructed with a partial sum of \eqref{eq:svd}. This reduces the dimensionality of the amplification mechanisms in \eqref{eq:resolvent}, only keeping the most significant contributions.

\section{General Features of Rotating Plane Couette Flow}\label{sec:rpcf}
We now introduce RPCF as the main flow configuration being used as a test case for the variational optimisation methodology. RPCF is defined on the same domain as the general planar wall-bounded flow in section~\ref{sec:variational-solver} with the same Cartesian coordinate system. The channel has a height of $2h$, the top and bottom walls move in opposite directions at a speed of $U_w$,  with the addition that the domain is rotated about the spanwise direction at a rotational rate of $\Omega$. The (non-dimensional) governing equations for this flow are given by
\begin{subequations}
    \begin{align}
    \pdv{\bm{u}}{t}&=-\left(\bm{u}\vdot\grad\right)\bm{u}-\grad{p}+\frac{1}{\mathit{Re}}\bm{\Delta}\bm{u}-\mathit{Ro}\left(\bm{e}_z\cross\bm{u}\right),\label{eq:ns-mom} \\
    \div{\bm{u}}&=0,
\end{align}
\label{eq:ns}%
\end{subequations}
 where the velocity is subject to the same boundary conditions as the flow described at the beginning of section~\ref{sec:methodology}. The flow is characterised by two non-dimensional parameters, the Reynolds number $\mathit{Re}$ as defined in section~\ref{sec:methodology} as well as the rotation number $\mathit{Ro}$
\begin{equation}
    \mathit{Ro}=\frac{2\Omega h}{U_w}.
\end{equation}
The rotation number is the ratio between the characteristic rotation and inertial forces of the system. In this work, only the rotation direction aligned with the shear is considered. This is known as anti-cyclonic rotation and acts to produce linear instabilities at finite Reynolds numbers \citep{tsukahara2010}, as opposed to standard plane Couette flow that is linearly stable for all finite Reynolds numbers \citep{daviaud1992}.  The most interesting parameter regime of the rotation number in this study is $0<\mathit{Ro}<1$, as linear instabilities in the flow are present in this range. For any value outside this range for $\mathit{Ro}$, the flow is actually made more linearly stable due to the presence of system rotation \citep{lezius1976,nagata2007}.



In this work we make an additional assumption that the flow is streamwise-independent ($\partial/\partial x=0$) reducing it to a 2-Dimensional 3-Component (2D3C) formulation. The primary reason to use a 2D3C formulation is to reduce the dimensionality of the problem making the solutions easier to find while retaining some of the key features of the original dynamics. This constraint in fact makes the flow exactly analogous to 2D Rayleigh-Benard convection as shown in \citet{eckhardt2020}. The analogy between rotating shear flows and convective thermal flows has been known for some decades \citep{chandrasekhar1961,yih1965}, and has been used to describe Taylor-Couette flows in turbulent regimes in relation to the Rayleigh Benard convection \citep{eckhardt2007}. The nature of some steady and streamwise-independent solutions to the RPCF at low Reynolds numbers (before bifurcating into more complex structures) is discussed in \citet{nagata2013} and \citet{nagata2021}. The streamwise-independence means that the domain size is given solely by the half-channel height $h$ and the spanwise length $L_z$, with the aspect ratio defined as the ratio $\gamma=L_z/h$. In this work an aspect ratio of $\gamma=4$ is used exclusively.

The flow starts with the stable laminar solution  $\bm{u}(y)=y\bm{e}_x$. Linear stability analysis shows that the boundary of stability for the laminar flow has the following relationship
\begin{equation}
    \mathit{Re}_{\text{crit}}=\sqrt{\frac{107}{\mathit{Ro}\left(1-\mathit{Ro}\right)}},
    \label{eq:laminar-bifurcation}
\end{equation}
given in \citet{lezius1976}. This initial bifurcation leads to a streamwise independent flow, and so it is also a valid relation for the 2D3C case. This implies that the laminar flow is most unstable for $\mathit{Ro}=0.5$, which is the rotation number of choice for this work. At $\mathit{Re}_\text{crit}\approx20.7$ the laminar  solution bifurcates and becomes unstable. A new stable equilibrium solution is born out of this bifurcation which has the characteristic streamwise-independent rolls that are present throughout the total set of regimes accessible by varying $\mathit{Re}$.

\begin{figure}
    \centering
    \includegraphics{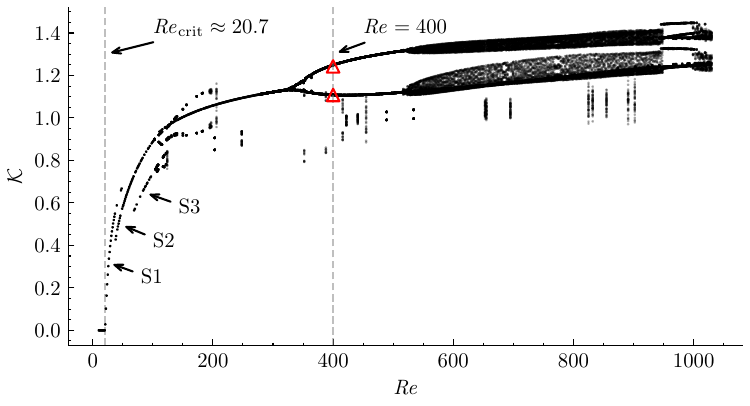}
    \caption{Bifurcation diagram of RPCF over a range of Reynolds number and $\mathit{Ro}=0.5$ showing the transition from the stable laminar solution to turbulent flow. Also plotted are the corresponding kinetic energy extrema of the periodic solution obtained from optimisation at $\mathit{Re}=400$ in section~\ref{sec:periodic}, denoted with red triangles.}
    \label{fig:dns-bifurcation}
\end{figure}

We characterise the route to turbulence by analysing time series extrema of the kinetic energy from DNS data over a range of Reynolds numbers at $\mathit{Ro}=0.5$, where the kinetic energy is defined as $\mathcal{K}(t)=\frac{1}{2}\norm{\bm{u}(t)}_{\Omega}$. All DNS results in this work were obtained using a custom solver \citep{lasagna2016} which employs a vorticity–streamfunction formulation, details of which are provided in appendix~\ref{app:dns-solver}. The results are shown in figure~\ref{fig:dns-bifurcation}. The purpose of this analysis is to provide the necessary context for interpreting the results from the variational optimiser. A comprehensive study of the transitionary behaviour of 2D3C RPCF lies beyond the present scope. For each Reynolds number, the DNS was initialised with a new random initial condition and then integrated until it converged to a final attractor, typically requiring thousands of time units. This procedure means that slightly different solutions may be obtained for different runs, depending on the initial condition.

\begin{figure}
    \centering
    \includegraphics{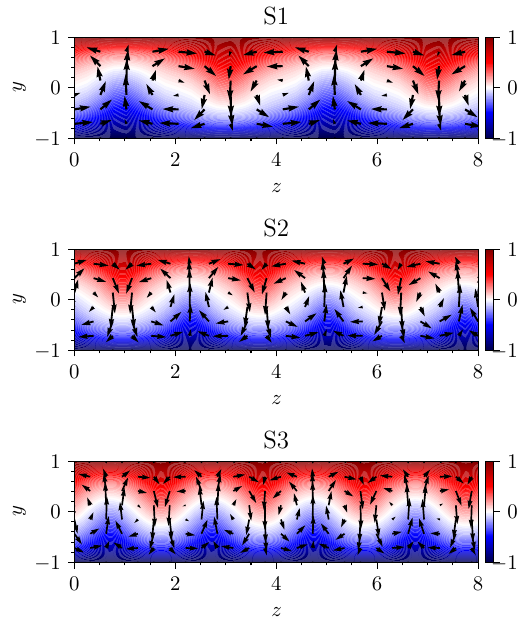}
    \caption{Equilibrium solutions labelled in figure~\ref{fig:dns-bifurcation} at  $\mathit{Re}=50$.  Contours represent the streamwise velocity and vectors represent the wall-normal and spanwise velocities.}
    \label{fig:dns-branches-snapshots}
\end{figure}

The initial bifurcation from the laminar  solution occurs at the expected value $\mathit{Re}\approx20.7$, as predicted from \eqref{eq:laminar-bifurcation}. The kinetic energy of the resulting stable equilibrium grows rapidly as $\mathit{Re}$ increases. Notably, instead of a single equilibrium branch being traced continuously, the DNS often settles on different attractors. These correspond to distinct stable equilibrium solutions, examples of which are labelled in figure~\ref{fig:dns-bifurcation}. At $\mathit{Re}=50$, three such branches, S1–S3, are shown in figure~\ref{fig:dns-branches-snapshots}. They differ primarily in the number of streamwise rolls and exhibit kinetic energies of approximately $\mathcal{K}=0.65$, $0.58$, and $0.39$ for S1, S2, and S3 respectively. The denser the rolls, the lower the kinetic energy, owing to higher dissipation at fixed Reynolds number. As $\mathit{Re}$ increases, denser rolling structures are favoured, likely because sparser structures cannot sustain the required dissipation rates. All of these equilibria are linearly stable at $\mathit{Re}=50$, since the DNS converges to them provided the initial condition lies within their basin of attraction. However, branches such as S3 are less commonly observed at this Reynolds number, likely because their basins of attraction are very small. 

As $\mathit{Re}$ increases further, each of these equilibria undergoes bifurcations to periodic solutions, sustained only over limited parameter ranges before seemingly losing stability and no longer acting as an attractor for the DNS. Around $\mathit{Re}=200$, the S1 branch again emerges as the dominant equilibrium, before bifurcating near $\mathit{Re}=350$ to a periodic orbit sustained up to a little above $\mathit{Re}=500$. This orbit retains the large-scale streamwise roll structure of its parent equilibrium. Beyond $\mathit{Re}=500$, the periodicity becomes increasingly irregular, with large amplitude fluctuations and a more broadband spectrum, indicating transition to turbulence. Finally, at the largest Reynolds numbers displayed in figure~\ref{fig:dns-bifurcation} the flow bifurcates again into a new apparently periodic solution before again transitioning to a chaotic evolution. These remaining bifurcations have not been investigated in any substantial way here.

In addition to the main branches, figure~\ref{fig:dns-bifurcation} also shows intermittent deviations that appear only over very narrow ranges of Reynolds number. These most likely arise because each DNS run was initialised with a different random condition, leading to variations in the transient dynamics. The precise origin of these deviations has not been investigated here, but they could correspond either to additional stable branches with small basins of attraction, which are only rarely observed, or to trajectories becoming temporarily trapped in transient regions of state space before converging to the dominant attractor. A more systematic exploration of state space would be required to determine their exact nature.

\section{Exact Solutions Found for RPCF}\label{sec:solutions}
In this section the focus is a demonstration of the projected variational optimisation methodology applied to finding exact nonlinear solutions to the Navier-Stokes equations for RPCF. 

\subsection{Implementation Details}\label{sec:numerics}
This section discusses the numerics and programmatic strategies used to implement the procedure depicted in figure~\ref{fig:optimisation-loop}. All numerics are implemented in the Julia programming language, the majority of the code is available at \href{https://github.com/The-ReSolver}{The-ReSolver}. A single spatially and temporally extended scalar field is represented as a 3D array, with each dimension representing the wall-normal, spanwise, and time directions, respectively. The number of wall-normal points, spanwise points, and temporal samples is denoted by $N_y$, $N_z$, and $N_t$, respectively. The actual array being stored for most computations, $\bm{u}_{\bm{k}}(y)$, has been transformed using a real FFT and thus has dimensions of $N_y\times(\lfloor N_z/2\rfloor+1)\times N_t$  for each of the three vector components, where $\lfloor\cdot\rfloor$ denotes the integer floor operation. The spanwise and time directions are transformed using FFTW \citep{fftw2005}, and so the discretisations in these directions are necessarily uniform. In this work the wall-normal points are also uniformly distributed to allow for easy interoperability between the discretisations used by the  DNS time-stepping code and Julia optimiser code.  The resolvent modes are represented by a 4D array. At each frequency tuple for the streamwise independent flow, $\bm{k}=(k_z,\,k_t)$, there are a set of $M\in\mathbb{N}$ modes containing $3N_y$ elements which are required to be stored in memory. Here where $M$ is the finite number of resolvent modes used in the truncated version of the sums \eqref{eq:velocity-expansion} and \eqref{eq:residual-expansion}. This leads to a array with dimensions of $3N_y\times M\times(\lfloor N_z/2\rfloor+1)\times N_t$. The modal coefficients $a_{\bm{k}m}$ (and $s_{\bm{k}m}$) are represented very similarly to the spectral arrays, only with the array dimension corresponding to the wall-normal direction now representing the modal index $m$, with a length $M$ for the number of modes used for the projection in \eqref{eq:velocity-projection} and \eqref{eq:residual-projection}. Thus, these arrays have a size of $M\times(\lfloor N_z/2\rfloor+1)\times N_t$. Since $M\sim N_y$ when the goal is to find exact solutions, the memory footprint of the coefficient arrays $a_{\bm{k}m}$ and $s_{\bm{k}m}$ is similar to the spectral arrays $\bm{u}_{\bm{k}}(y)$. The largest component of the memory requirements for finding exact solutions comes from the modes themselves, where all dimensions of the array grow with Reynolds number. This problem is highly parallelisable due to the global representation, which would be a necessary step to implementing these methods for fully 3D problems or for higher Reynolds numbers.

Derivatives in the spanwise and time directions can be computed using standard spectral methods. Derivatives in the wall-normal direction are computed using finite difference methods, generally of second order accuracy to match as closely with the numerics of the DNS solver. The finite difference stencils are derived using the custom package \href{https://github.com/The-ReSolver/FDGrids.jl}{FDGrids.jl}. To compute the projection in \eqref{eq:velocity-projection} and \eqref{eq:residual-projection} it is necessary to compute integrals over all the directions given the definition of the  inner product in \eqref{eq:inner-product}. In the spanwise and time directions this can be done by using Parseval's theorem, converting the integral over these directions to sums of the coefficients over the corresponding wavenumbers. The wall-normal integrals are computed using the method of undetermined coefficients for quadratures \citep{dahlquist2008}. The nonlinear terms in the computations are computed using a pseudo-spectral method, using a $3/2$ padding rule in the spanwise and time directions to avoid aliasing errors.

When constructing the modes $\bm{\psi}_{\bm{k}m}$ using resolvent analysis, the base flow used is the laminar solution  $\bm{u}_b(y)=y\bm{e}_x$. This choice is made for its computational simplicity and numerical robustness; the laminar base flow is well-defined, analytically simple, and guarantees a tractable, well-conditioned singular value decomposition for all wavenumbers $\bm{k}\in\mathbb{Z}^2$ (including $\bm{k}=0$). Consequently, it provides a reliably calculable set of modes $\bm{\psi}_{\bm{k}m}$ without requiring a priori knowledge of the flow or introducing the numerical complications associated with a turbulent mean profile.  We have observed that the difference in the modes generated using the laminar profile and turbulent mean do not have significant qualitative differences, thus the optimisation is not greatly affected by this choice as long as a sufficient number are used.  To compute the resolvent modes the resolvent operator in \eqref{eq:resolvent} has to be discretised. That is, for each wavenumber tuple $\bm{k}$, the operator $\bm{R}_{\bm{k}}$ has to be evaluated for the given discrete representation of the wall-normal derivative $\pdv*{y}$. This process is done using primitive variables ($\bm{u}$ and $p$) as depicted in \citet{symon2018}. The resolvent takes the form as a matrix, augmented with the incompressibility constraint and the pressure gradient term, for each $\bm{k}$, which is then inverted. Before performing the SVD the resulting matrix has to be transformed such that it obeys the correct orthogonality condition defined in \eqref{eq:mode-orthonormal} \citep{sipp2013,mckeon2019}. To accomplish this the quadrature weights used to evaluate the wall-normal integrals are Cholesky decomposed, which provides a weighting matrix that ensures the resulting SVD will be orthogonal with respect to the correct inner-product space. The divide and conquer eigenvalue algorithm is used to perform the SVD.

We are free to choose the optimisation algorithm used for solving the optimisation problem \eqref{eq:opt-prob}, since any gradient-based method will work.  Unless otherwise stated, the Limited-memory BFGS (L-BFGS) algorithm is used. L-BFGS is a quasi-Newton method, incorporating approximate Hessian information into each iteration, which significantly improves the convergence properties of the optimisation near minima compared to gradient descent  \citep{liu1989,fletcher2000chapter2}. This is particularly important here because the optimisation problem is non-convex, with many possible solutions potentially very close together. The problem is also generally quite poorly conditioned, a topic that will be further expanded upon in section~\ref{sec:conditioning}. 
L-BFGS's ability to approximate the local curvature of the solution reduces the effect of this poor conditioning. Further discussions on each algorithm can be found in \citet{nocedal2006}. The optimisation is performed using the \href{https://github.com/JuliaNLSolvers/Optim.jl}{Optim.jl} package \citep{optim.jl}. Any algorithm used here is also coupled with the Hager-Zhang line search algorithm as described in \citet{hager2005-linesearch}. This further improves convergence rates, albeit at the possibility of reducing the robustness of the convergence. In practice, however, this is not observed to be an issue. The threshold for convergence is defined as $\mathcal{R}[\bm{u}]<10^{-12}$.

\subsection{Equilibria}\label{sec:equilibria}
\begin{figure}
    \centering
    \includegraphics{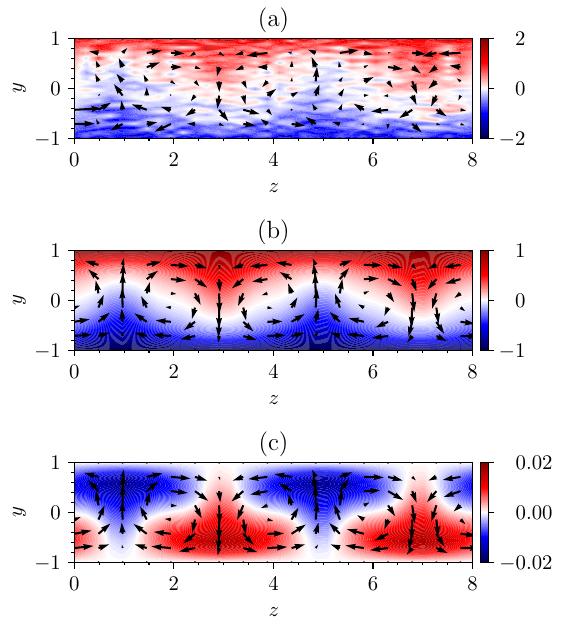}
    \caption{Snapshots of the flows before and after the optimisation at $\mathit{Re}=50$ and $\mathit{Ro}=0.5$, along with the solution obtained from DNS. Panel (a) shows the initial flow used for the optimisation, obtained by perturbing the stable solution obtained from DNS at the same Reynolds and rotation numbers. Panel (b) shows the result of the optimisation with a residual of $\mathcal{R}<10^{-12}$, and panel (c) is difference between the optimisation result and the DNS solution.   Contours represent the streamwise velocity and vectors represent the wall-normal and spanwise velocities.}
    \label{fig:EQ1-init-final-snapshot}
\end{figure}

The Reynolds number is set to $\mathit{Re}=50$ and rotation number $\mathit{Ro}=0.5$ for the equilibrium solutions. To begin, the optimiser is validated by observing its ability to reconstruct a known solution to the flow. This was obtained using the custom DNS solver, with a grid discretisation of $N_y=64$ on a uniform grid, and $N_z=33$ corresponding to $17$ spanwise Fourier modes. A set of $M=64$ resolvent modes $\bm{\psi}_{\bm{k}m}$ are used for this optimisation. To obtain the initial guess for the start of the optimisation, S1 from DNS in figure~\ref{fig:dns-branches-snapshots} is projected onto the resolvent basis and then perturbed with random Gaussian white noise at each coefficient $a_{\bm{k}m}$. This results in the highly disordered field in figure~\ref{fig:EQ1-init-final-snapshot}(a). This perturbed flow is then optimised to try to recover the original solution. The result of this optimisation is shown in figure~\ref{fig:EQ1-init-final-snapshot}(b). This can be compared to S1 in figure~\ref{fig:dns-branches-snapshots}, where the equilibria from the DNS and optimisation are qualitatively indistinguishable. Figure~\ref{fig:EQ1-init-final-snapshot}(c) shows the difference between the solution obtained from optimisation and from the DNS. The magnitude of the difference is a couple of orders magnitude smaller than that of the actual solution.  The discrepancy likely arises from differences between the optimisation framework and the DNS formulation, rather than from the optimisation algorithm itself. Specifically, the optimisation is based on a discretisation of the continuous adjoint operator. It is known the continuous adjoint of the linearised operator being studied, when discretised, does not exactly correspond to the adjoint of the discretised linearised operator, producing errors that only disappear as the resolution of the discretisation increases \citep{nadarajah2000,hekmat2016,gao2017}. A limited convergence study was performed in which the finite difference stencil widths, as well as the wall-normal and spanwise resolutions, were systematically varied. Although both the Julia optimiser and DNS solver exhibited internal convergence, the relative discrepancy between them did not decrease monotonically with refinement, indicating that it is not solely attributable to insufficient spatial resolution.

This validates the optimiser possesses minima that correspond to solutions of \eqref{eq:ns}.

\begin{figure}
    \centering
    \includegraphics{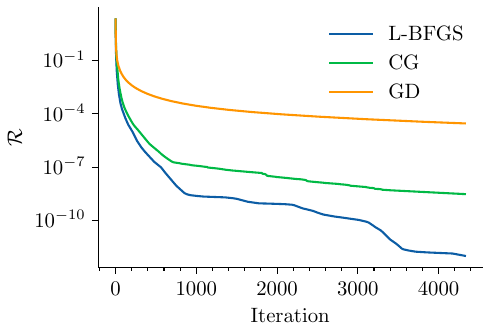}
    \caption{Residual trace for the optimisation of the initial flow given in panel (a) of figure~\ref{fig:EQ1-init-final-snapshot}, using Gradient Descent (GD), Conjugate Gradient (CG), and L-BFGS optimisation algorithms. All solutions converge towards the solution obtained in panel (b) of figure~\ref{fig:EQ1-init-final-snapshot}.}
    \label{fig:EQ1-trace}
\end{figure}

For comparison, the same initial condition was also optimised using the gradient descent and conjugate gradient algorithms (both still using Hager-Zhang line searches). The traces for each algorithm applied to the initial condition in figure~\ref{fig:EQ1-init-final-snapshot}(a) are shown in figure~\ref{fig:EQ1-trace}. Clearly L-BFGS performs the best, achieving the residual of $\mathcal{R}=10^{-12}$ after a little over $4{,}000$ iterations. Gradient descent displays an initial large decrease in residual which quickly decreases as the convergence rate slows. This final and slow phase of the optimisation is a result of the optimiser approaching the minimum along the direction associated with the slowest growth direction in $\mathcal{R}$. Conjugate gradient also outperforms gradient descent, achieving a smaller residual for the same number of iterations. However, after the initial large decrease in the residual at the beginning of the optimisation, the convergence rate slows, approaching a similar speed as gradient descent. In addition, the iteration time of conjugate gradient is observed to be larger than L-BFGS.

\begin{figure}
    \centering
    \includegraphics{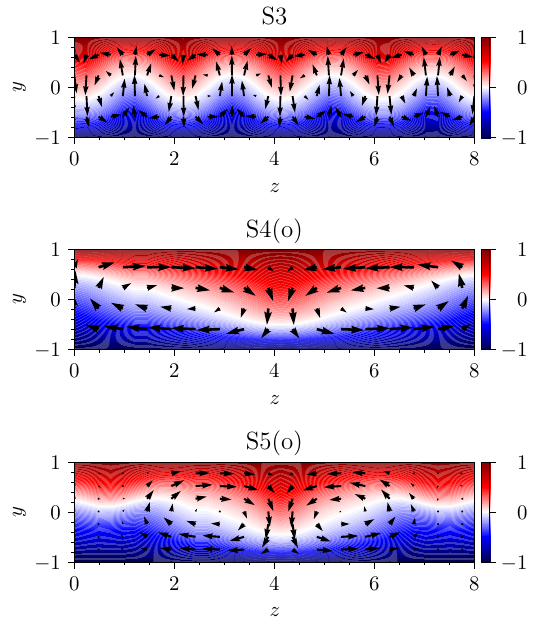}
    \caption{Final snapshots of the solutions obtained by optimising from various synthetic initial flow fields.  Contours represent the streamwise velocity and vectors represent the wall-normal and spanwise velocities.}
    \label{fig:unstable-equilibrium-snapshots}
\end{figure}

To investigate the robustness of the optimiser, it was initialised with initial conditions exciting certain spanwise wavenumbers with the intent of finding multiple new equilibrium solutions that are not observed in DNS at this Reynolds number. The same set of resolvent modes used for the previous optimisation are utilised here. The results of these optimisations are shown in figure~\ref{fig:unstable-equilibrium-snapshots}. Figure~\ref{fig:unstable-equilibrium-snapshots}(a) was initialised with the first $3$\textsuperscript{rd} to $5$\textsuperscript{th} modal coefficients $a_{\bm{k}m}$ excited with random values at the spanwise wavenumber of $k_z=4$, corresponding to streamwise rolls with half the wavelength of S1, along with as the $3$\textsuperscript{rd} mode at $k_z=0$, with the rest of the coefficients left as zero. This synthetically constructed initial flow converges to S3, one of the stable equilibria shown in figure~\ref{fig:dns-branches-snapshots}. Next, when the first $5$ modal coefficients are excited randomly for the spanwise wavenumber $k_z=1$, the solution in figure~\ref{fig:unstable-equilibrium-snapshots}(b) is found, called S4(o). The addition of ``(o)'' in the name given to the equilibrium is in reference to the fact that it has been obtained from optimisation and is not a solution observed from the DNS, a result of them likely being linearly unstable. This solution again has the streamwise roll expected from the other equilibria solutions, but now with twice the spanwise wavelength. The highly symmetric structures of S3 and S4(o) closely mirror the solution of  S1 and are ultimately expressing the same dynamics. To obtain a slightly more abnormal solution, the flow was initialised by exciting the first $8$ resolvent modes at the zeroth and first ($k_z=0,1$) spanwise wavenumbers with random values. The result of optimising such an initial condition is shown in figure~\ref{fig:unstable-equilibrium-snapshots}(c), called S5(o). S5(o) shows a streamwise roll pattern with a width larger than in S1, roughly 3 spanwise units in length. This roll does not repeat over the remaining length of the domain since there is not sufficient room. Instead, the flow remains mostly stagnant in the remaining space, only expressing some very weak rolls that transports a small amount of the momentum from the top wall downwards.

\begin{figure}
    \centering
    \includegraphics{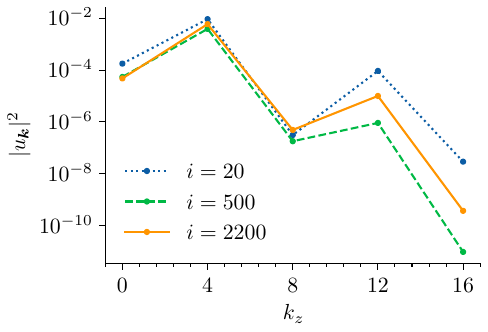}
    \caption{Spanwise power spectra of S3 of the solutions obtained at specific iterations of its optimisation, sampled at the channel  mid-plane} ($y=0$).
    \label{fig:EQ2-spectra-convergence}
\end{figure}

The spanwise power spectra of the intermediate flow over the duration of the optimisation leading to S3 are shown in figure~\ref{fig:EQ2-spectra-convergence}.  That is, the squared magnitude of the Fourier coefficients of the solution S3 along the spanwise direction sampled at the mid-plane $y=0$. For clarity, the wavenumbers that are orders of magnitude smaller than those shown are omitted. The omitted wavenumbers have negligible impact on the solution and are primarily a result of the spanwise domain size being large enough to contain multiple repetitions of the fundamental unit of the solution. The initial field used for the optimisation was excited at exactly one spanwise wavenumber, $k_z=4$. The optimiser initially spreads out the spectral content from the initial condition to the other wavenumbers as can be seen at iteration $i=20$. The large wavenumbers are then damped by viscous effects and the energy containing low wavenumbers are tuned towards the final solution by $i=500$. The remaining iterations are focused on finely tuning the higher wavenumbers while gradually decreasing the residual, until a balanced flow is achieved at $i=2{,}200$.  The effects of the individual terms in the gradient and how they affect the optimisation are further discussed in section~\ref{sec:conditioning}

 A preliminary natural continuation of equilibrium S1, implies that these equilibrium solutions do not exist at the larger Reynolds number regimes shown in section~\ref{sec:rpcf}. It is not known, however, whether these equilibrium solutions exist in the higher Reynolds number regimes. The only definitive way to answer this question to perform a rigorous continue the solutions to higher Reynolds numbers. With this comes the expected additional computational costs associated with the finer resolution required in more energetic regimes.

\subsection{Periodic Solutions}\label{sec:periodic}
The optimiser is now tasked with constructing a periodic solution at $\mathit{Re}=400$, a regime where RPCF displays a stable periodic motion as seen in figure~\ref{fig:dns-bifurcation}. The flow is discretised with $N_y=128$, $N_z=65$, corresponding to $33$ spanwise modes, and $N_t=35$ temporal modes. A set of $M=64$ resolvent modes for each frequency are used for the optimisation. The smaller number of resolvent modes is used primarily to reduce computational effort. As will be seen in this section, the reduced number of modes compared to the  degrees of freedom in the wall-normal direction has little effect on the accuracy of the final result. The initial guess for the optimiser is initialised in a similar way as in the first validation case in section~\ref{sec:equilibria}.  L-BFGS is still used for the optimisation along with a Hager-Zhang line search. The periodic DNS data at $\mathit{Re}=400$ is projected onto the resolvent modes. The resulting coefficients $a_{\bm{k}m}$ are then perturbed with random Gaussian white noise and the resulting noisy flow is optimised.

\begin{figure}
    \centering
    \includegraphics{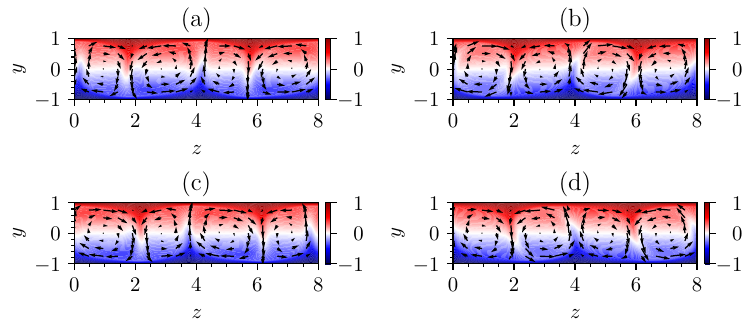}
    \caption{Snapshots of the periodic solution ($\mathcal{R}\approx5\times10^{-11}$) obtained for $\mathit{Re}=400$ with a period of $T\approx25.05$. Panel (a) shows $t=0$, panel (b) shows $t=T/4$, panel (c) shows $t=T/2$, and panel (d) shows $t=3T/4$.  Contours represent the streamwise velocity and vectors represent the wall-normal and spanwise velocities.}
    \label{fig:periodic-solution-snapshots}
\end{figure}

\begin{figure}
    \centering
    \includegraphics{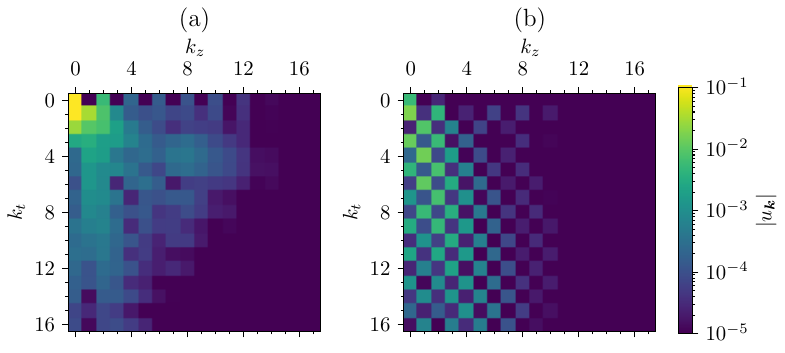}
    \caption{Spanwise and (positive) temporal power spectrum of the periodic solution in figure~\ref{fig:periodic-solution-snapshots} at $y\approx-0.86$ in panel (a) and $y\approx0$ in panel (b).}
    \label{fig:periodic-solution-spectrum}
\end{figure}

Figure~\ref{fig:periodic-solution-snapshots} shows a set of snapshots for the periodic solution obtained from optimisation at $\mathit{Re}=400$, sampled at points along its trajectory. The primary streamwise rolls are clearly present, and evolve in a wavy motion as consecutive vortices contract and expand. Figure~\ref{fig:periodic-solution-spectrum} shows the spectrum of the periodic solution sampled at the two wall-normal positions $y\approx-0.86$ and $y\approx0$. Each spectrum is a slice of the total spectrum such that only the positive temporal frequencies are shown, since the negative temporal frequencies are a reflection. An imbalance in the positive and negative temporal frequencies would manifest as travelling waves moving in the spanwise direction, which is not a feature of this solution. The general lack of spanwise motions in solutions to channel flows is a noted feature in \citet{cvitanovic2010}. A chequerboard pattern is observed in the spectrum near the centreline of the flow.  In the temporal spectrum, this is a result of the periodic solution being made up of a pre-periodic solution repeating itself under a symmetry transformation. Pre-periodic solutions are able to exist in this flow due to the presence of certain discrete and continuous symmetries \citep{gibson2008}, however, such symmetric solutions are not treated in this work. The pattern in the spanwise direction, present only near the centre-plane is simply a result of the strong streamwise rolls that dominate the flow at this Reynolds number. The most energetic mode is located at $\left(k_z,\,k_t\right)=\left(1,\,0\right)$ which is the mode that best fits the streamwise rolls that do not vary significantly with time. The spectrum at the location nearer the wall has a more continuous decay indicative of its less obvious spatial and temporal structure compared to the centre of the channel.

\begin{figure}
    \centering
    \includegraphics{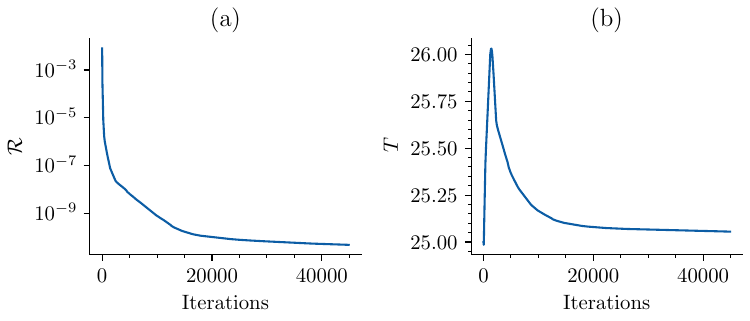}
    \caption{Traces of the global residual, panel (a), and the solution period, panel (b), of the periodic solution in figure~\ref{fig:periodic-solution-snapshots} over the duration of the optimisation.}
    \label{fig:periodic-solution-trace}
\end{figure}

Figure~\ref{fig:periodic-solution-trace} shows the trace of the global residual and period in panels (a) and (b), respectively. The final residual achieved is roughly $\mathcal{R}\approx5\times10^{-11}$. The convergence rate of the problem is considerably slower after the initial rapid drop in residual than that observed in figure~\ref{fig:EQ1-trace}. This slower convergence compared to that observed in figure~\ref{fig:EQ1-trace} can be explained by the increased  degrees of freedom of the problem, both in terms of the inclusion of an extra temporal dimension and in terms of the added resolution required at higher Reynolds numbers.  It is also likely that this particular solution at the Reynolds number of $\mathit{Re}=400$ is near marginal stability, however analysis of the stability of these solutions has not been undertaken in this work. The impact of marginal stability is discussed further in section~\ref{sec:conditioning}. As a final point, the maximal and minimal kinetic energies of the optimised periodic solution  are plotted on figure~\ref{fig:dns-bifurcation} (denoted with red triangles) with the corresponding kinetic energy extrema from DNS, which shows good agreement between the solutions from the two different numerical sources.




\subsection{Performance of Variational Optimisation}
The performance of the variational optimisation is best understood in comparison with the leading methods used to compute time-periodic solutions of nonlinear dynamical systems, namely the variational Newton flow method of \citet{lan2004} and the Newton–GMRES–hookstep shooting method \citep{viswanath2007,viswanath2009}. Both are Newton-based approaches and therefore require the solution of a linear system at each iteration. For the variational Newton flow method, the associated linear operator has dimension $N\times N$, where $N$ is the total number of degrees of freedom. For realistic fluid flow problems, this renders the method computationally intractable without any modification.

The Newton–GMRES–hookstep shooting method circumvents the explicit formation of this operator by modifying the Newton iteration with a Krylov subspace method, yielding a matrix-free implementation suitable for high-dimensional turbulent flows. Nevertheless, each Newton step typically requires tens to hundreds of matrix–vector products at least in order to solve the linear system to sufficient accuracy. In contrast, the variational optimisation method evaluates the gradient, and hence the update direction, using a single matrix–vector product per iteration. It is therefore significantly cheaper per iteration. Additionally, as demonstrated here and in \citet{farazmand2016,schneider2022,schneider2023}, the variational optimisation is also substantially more robust to the choice of initial guess. More formally, convergence to a minimum is guaranteed, whereas Newton-based methods may fail to converge unless the initial guess lies sufficiently close to a solution.

These advantages, however, come at the expense of linear convergence near a minimum, in contrast to the quadratic convergence attainable with Newton methods. Consequently, once the residual has been reduced below a moderate threshold, it becomes more efficient to switch to Newton–GMRES–hookstep in order to drive the residual to machine precision. This observation motivates the hybrid strategy proposed by \citet{farazmand2016}. In the present computations, the first $40{,}000$ iterations shown in figure~\ref{fig:periodic-solution-trace} required approximately three days of wall-time, with each iteration taking ${\sim}7$ seconds. However, the first $2{,}000$ iterations, sufficient to reduce the residual below $\mathcal{R}<10^{-7}$, required only around four hours.

In the specific formulation of the problem here, the memory cost of the algorithm is dominated by the storage of the modes $\bm{\psi}_{\bm{k}m}$. The memory consumed by these modes scales with $\mathcal{O}\left(MN\right)$ where $M$ is the total number of resolvent modes used, and $N$ here represents the combination of the spatial and temporal degrees of the freedom of the problem. For 2D3C RPCF this is given as $N=3N_y\left(\lfloor N_z/2\rfloor+1\right)N_t$. For reference, the modes used for the optimisation of the periodic solution shown in figure~\ref{fig:periodic-solution-snapshots} consumed ${\sim}430$MB. Using the Kolmogorov hypothesis it can be estimated that the number of grid points required in each direction scales as $\mathit{Re}^{3/4}$ \citep{pope2001}. Using this scaling, and assuming that the number of modes $M$ scales in a similar way to the grid points, it can be estimated that the required memory to store the modes at a fully turbulent Reynolds number of $\mathit{Re}=4000$ (up from $\mathit{Re}=400$) is ${\sim}420$GB. Clearly this is becomes a prohibitively expensive endeavour as the Reynolds number grows. The most practical solution is to distribute the data over multiple processes to reduce the local memory requirements. Further discussions on the performance of the variational optimiser in general can be found in \citet{schneider2022}. The time complexity of the algorithm is dominated by the Fourier transforms performed to compute the nonlinear terms in \eqref{eq:projected-gradient} and \eqref{eq:projected-residual}, scaling as $\mathcal{O}\left(N_yN_{\text{hom}}\log\left(N_{\text{hom}}\right)\right)$ where $N_{\text{hom}}=\left(\lfloor N_z/2+1\rfloor\right)N_t$ are the degrees of freedom of the homogenous spatial directions and time. This cannot be made any more efficient in time than its current form.

\section{Conditioning of the Optimisation}\label{sec:conditioning}
In the literature the convergence rate of variational optimisation algorithms slows considerably near a minimum, a problem governed by the local curvature of the residual function \citep{nocedal2006}. This curvature is characterized by the Hessian operator, which contains all second-derivative information at a point in optimisation space. The condition number of the Hessian measures the scale separation between the slowest and most rapid growth rates experienced by the objective when moving away from the minimum. A lower condition number implies improved convergence rates for gradient-based optimisation algorithms. A condition number of one signifies a perfectly quadratic local neighbourhood, while larger values indicate a long, narrow ``valley'', which impedes convergence. The goal of this section is to link this conditioning directly to the underlying flow dynamics and to demonstrate that the resolvent modes provide a basis optimally constructed to improve the conditioning of the optimisation problem.

The first step in the analysis is to derive a closed-form expression for the Hessian operator of the residual as defined in \eqref{eq:opt-prob}. To do this, first consider a minimum $\mathcal{R}[\bm{u}^*]=0$, with $\bm{u}^*$ denoting the flow field that solves the Navier-Stokes equations. Since $\bm{u}^*$ is a minimum of \eqref{eq:opt-prob} then we also know that $\delta\mathcal{R}/\delta\bm{u}^*=0$. Assume that $\bm{u}^*$ is an equilibrium. This does not change the content of the discussions that follow, only serving to simplify the mathematics by removing the need to use Floquet analysis. Adding an infinitesimal (potentially unsteady) perturbation to $\bm{u}^*$ in the direction $\bm{v}$, where $\bm{v}$ is incompressible and obeys a set of homogeneous boundary conditions similar to the local residual in \eqref{eq:adjoint-constraints}, and expanding in terms of a Taylor series, only retaining the second-order term, leads to the following relation
\begin{equation}
    \innprod{\bm{v}}{\bm{H}\bm{v}}_{\Omega_t}=\norm{\pdv{\bm{v}}{t}-\mathbb{P}\mathcal{L}_{\bm{u}^*}\bm{v}}_{\Omega_t}^2,
    \label{eq:hessian-expression}
\end{equation}
where $\bm{H}$ denotes the Hessian operator, $\mathcal{L}_{\bm{u}^*}$ denotes the linearised Navier-Stokes operator evaluated at the minimum $\bm{u}^*$, and $\mathbb{P}$ is the Leray projector  from section~\ref{sec:resolvent}. The inclusion of the Leray projector is as a notational short-hand to avoid the added difficulty in having to explicitly account for the pressure gradient term in the Navier-Stokes equations. The detailed derivation of this equation is given in Appendix~\ref{app:hessian-derivation}. Let $\bm{v}$ be a unimodal perturbation to that minimum, i.e. let $\bm{v}=\bm{v}_0e^{i\omega t}+\text{c.c.}$ for some arbitrary frequency where $\omega\in\mathbb{R}$ and some steady flow field $\bm{v}_0$, where c.c. denotes the complex conjugate. This perturbation can be substituted into \eqref{eq:hessian-expression}, effectively performing a Fourier transform, which gives
\begin{equation}
    \innprod{\bm{v}_0}{\bm{H}_\omega\bm{v}_0}_{\Omega}=\norm{\left(i\omega\bm{I}-\mathbb{P}\mathcal{L}_{\bm{u}^*}\right)\bm{v}_0}_{\Omega}^2.
    \label{eq:equilibrium-hessian}
\end{equation}
The Hessian operator is self-adjoint, as is apparent from \eqref{eq:hessian-expression}, which means it is also normal. As such, the eigenvalues obey a strict ordering
\begin{equation}
    0\leq\mu_{\omega1}\leq\mu_{\omega2}\leq\cdots\leq\mu_{\omega n}\leq\cdots.
\end{equation}
This sequence of eigenvalues is unbounded since the Hessian itself is an unbounded operator. This is equivalent to saying that there exists no positive number, $L$, such that $\mu_{\omega n}\leq L$ for all $n\in\mathbb{N}$. In practical terms, when the governing equations are discretised in space this leads to a new discretised Hessian operator, denoted as $\hat{\bm{H}}_\omega$, which does have a bounded spectrum. The relationship between the spectra of $\bm{H}_\omega$ and its discretised counterpart $\hat{\bm{H}}_\omega$ depends on the particular discretisation used and is not the focus of the present analysis. Instead, the focus is that of the largest eigenvalue of the Hessian that can be resolved from $\hat{\bm{H}}_\omega$. Defining this most extreme element of the spectrum as $\mu_{\omega N}$ it is possible to then define a condition number of the (discrete) Hessian: $\kappa(\hat{\bm{H}}_\omega)=\mu_{\omega N}/\mu_{\omega1}$.

It has been documented that the convergence rate of variational optimisers degrades near marginally stable solutions or bifurcation points \citep{farazmand2016,lakoba2007}. This phenomenon can be understood by considering the linear stability of the equilibrium solution $\bm{u}^*$, governed by the spectrum of $\mathcal{L}_{\bm{u}^*}$. Inspecting the derived expression for the Hessian in \eqref{eq:equilibrium-hessian}, it is clear that in the limit of any eigenvalue of $\mathcal{L}_{\bm{u}^*}$ approaching the imaginary axis (neutral stability), the action of the operator $i\omega\bm{I}-\mathcal{L}_{\bm{u}^*}$ becomes arbitrarily small near the frequency $\omega$ where that eigenvalue crosses. This implies that any perturbation $\bm{v}_0$ aligned with this marginally stable direction has a vanishingly small action from the Hessian, meaning the residual grows very slowly in this direction. In relation to the spectrum of $\bm{H}_\omega$, this is equivalent to the smallest Hessian eigenvalue $\mu_{\omega1}$ approaching zero, which forces the condition number $\kappa(\hat{\bm{H}}_\omega)\to\infty$. This mathematical insight is the basis for acceleration methods that assume the optimiser is traversing along the most marginally stable mode \citep{schneider2022,schneider2023}, and several techniques have been devised to try and remove these problematic directions \citep{yang2007}, although they have not been applied to fluid dynamics problems.

\begin{figure}
    \centering
    \includegraphics{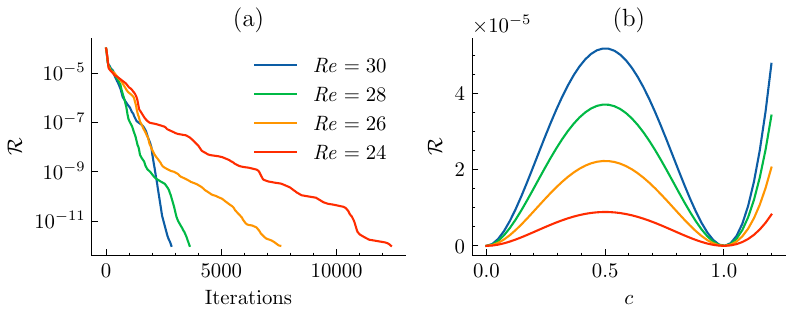}
    \caption{Residuals from optimisations of S1 at Reynolds numbers of $\mathit{Re}=30,\,28,\,26,\,24$. Panel (a) shows the residuals achieved during an optimisation of the equilibrium at each Reynolds number, and panel (b) shows the residual values  of the linearly interpolated velocity field between the final equilibrium solution and the laminar solution, $\bm{u}(c)=\bm{u}_{\text{lam}}+c(\bm{u}_{\text{EQ}}-\bm{u}_{\text{lam}})$.}
    \label{fig:bifurcation-convergence}
\end{figure}

This behaviour is demonstrated in figure~\ref{fig:bifurcation-convergence}(a), which shows the global residual over the course of a series of optimisation at different Reynolds numbers. As mentioned in section~\ref{sec:rpcf} at $\mathit{Re}_crit\approx20.7$ a bifurcation occurs where a new stable equilibrium solution is born out of the laminar  solution. Thus, as the Reynolds number gets closer to this bifurcation, the stable equilibrium gradually becomes more neutrally stable, leading to degrading convergence rates.  Figure~\ref{fig:bifurcation-convergence}(b) plots the global residual of the linearly interpolated flow field between the laminar  solution and the equilibrium solution obtained from the optimisation at the corresponding Reynolds number, given by the expression $\bm{u}(c)=\bm{u}_{\text{lam}}+c\left(\bm{u}_{\text{EQ}}-\bm{u}_{\text{lam}}\right)$, where $c\in[0,\,1]$. The variation of global residual between the laminar solution and the equilibrium solution decreases as the Reynolds number approaches $\mathit{Re}_crit$. This demonstrates the emergence of a near neutral direction for the optimisation. The rate at which the convergence degrades as the Reynolds number approaches neutrality depends on the type of bifurcation that is occurring and how fast the particular eigenvalue that is switching stability approaches the imaginary axis.

The relationship in \eqref{eq:equilibrium-hessian} can be rearranged to provide a closed-form expression for the  continuous Hessian operator
\begin{equation}
    \bm{H}_\omega\bm{v}_0=\left(i\omega\bm{I}-\mathbb{P}\mathcal{L}_{\bm{u}^*}\right)^+\left(i\omega\bm{I}-\mathbb{P}\mathcal{L}_{\bm{u}^*}\right)\bm{v}_0,
    \label{eq:hessian-product}
\end{equation}
where $(\cdot)^+$ denotes the adjoint of an operator.  Taking the definition of the resolvent operator $\bm{R}_\omega=(i\omega\bm{I}-\mathbb{P}\mathcal{L}_{\bm{u}^*})^{-1}\mathbb{P}$ from \eqref{eq:resolvent}, and applying the SVD defined in \eqref{eq:svd} the non-normal growth mechanisms of $\mathcal{L}_{\bm{u}^*}$ can be related to the spectrum of $\bm{H}_\omega$
\begin{equation}
    \bm{H}_\omega\bm{v}_0=\sum_{m=1}^\infty\sigma_{\omega m}^{-2}\bm{\psi}_{\omega m}\innprod{\bm{\psi}_{\omega m}}{\bm{v}_0}_{\Omega}.
    \label{eq:hessian-expansion}
\end{equation}
Equation~\eqref{eq:hessian-expansion} represents an eigendecomposition of the Hessian, directly linking its spectrum to the resolvent of $\mathcal{L}_{\bm{u}^*}$. The eigenvalues of $\bm{H}_\omega$ are the squared inverse of the resolvent's singular values ($\mu_{\omega m} = \sigma_{\omega m}^{-2}$), and their corresponding eigenmodes are identical to the resolvent response modes. Therefore, the condition number  of the discretised operator $\kappa(\hat{\bm{H}}_\omega)$ is primarily governed by the smallest retained singular value when the expansion is discretised. This relationship provides an optimal strategy for preconditioning since by truncating the resolvent expansion to remove the modes with the smallest singular values, one selectively eliminates the largest, most problematic eigenvalues from the Hessian. This optimally reduces the condition number, as it precisely targets and removes the directions associated with the strongest growth in the residual. In \citet{mons2021} the authors attempted to derive an methodology to determine optimal sensor placement for data assimilation. Their formulation lead to a similar relationship to \eqref{eq:hessian-product} relating the Hessian operator of their problem to the resolvent of their linearised dynamics. The fact that this relationship appears in both this work and \citet{mons2021}, despite the different objectives, is indicative of the role the resolvent has in modelling the higher order sensitivities in these types of optimisation problems.

A key limitation is that these optimal resolvent modes depend on the linearised operator $\mathcal{L}_{\bm{u}^*}$ evaluated at  an a priori unknown minimum. Thus, to be able to derive the modes used in \eqref{eq:hessian-expansion} the solution has to be known before the optimisation. Therefore it appears that this does not provide a useful method for improving convergence rates. We posit, however, that the higher-order modes exhibit universal structure, independent of the base flow, due to the dominance of viscous dissipation at small scales. This is favourable, as \eqref{eq:hessian-expansion} indicates that rejecting these higher-order modes is the principal method for improving the Hessian's condition number. It should be noted that the sensitivity of resolvent modes to the base flow used for the linearisation remains poorly characterised, beyond the established dependence of the singular values in \citet{brandt2011}.

\begin{figure}
    \centering
    \includegraphics{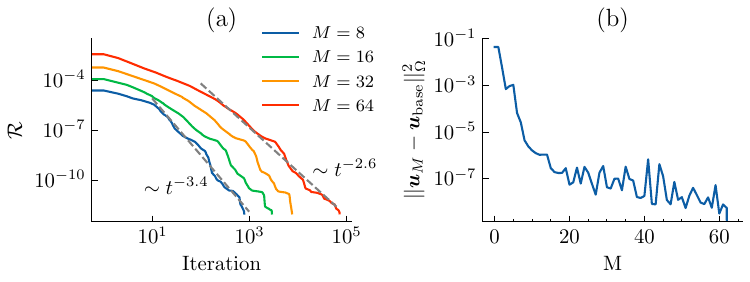}
    \caption{Panel (a): global residual traces for the optimisation of a perturbed S3 solution, performed using $M=8$, $16$, $32$, and $64$ resolvent modes, all  with the same set of initial coefficients, using the L-BFGS algorithm. Panel (b): Accuracy of the resulting solutions found by the optimiser relative to the ``base'' case obtained for $64$ resolvent modes, plotted against the number of modes used for the projection, each solution being converged such that $\mathcal{R}<10^{-12}$.}
    \label{fig:traces-vs-modes-EQ2}
\end{figure}

To demonstrate that rejecting higher order resolvent modes improves the convergence rate of the optimisation, we perform a series of optimisations starting from a perturbed state close to S3 from section~\ref{sec:rpcf} and \ref{sec:equilibria} at $\mathit{Re}=50$. The optimisation is performed using progressively fewer resolvent modes to generate the low-order model. The modes used are the same as those used in section~\ref{sec:equilibria}, derived by linearising about the laminar base flow. The results of these optimisations are shown in figure~\ref{fig:traces-vs-modes-EQ2}.

Figure~\ref{fig:traces-vs-modes-EQ2}(a) shows the global residual traces for the perturbed field using $8$, $16$, $32$, and $64$ resolvent modes. All the residuals reduce at a high rate initially, indicating the power of the variational optimiser to quickly seek out the solution primarily by modifying the large-scale structures.  The initially fast decrease, however, eventually gives way to a slower convergence rate  governed by a power-law, as is typically observed in the literature \citep{schneider2022,schneider2023}. The primary point to note from figure~\ref{fig:traces-vs-modes-EQ2}(a) is that both the degree of the eventual slowdown in convergence rate and the iteration count at which it occurs are linked to the  degree of truncation of the sums in \eqref{eq:velocity-expansion} and \eqref{eq:residual-expansion}, i.e. how many dimensions the resulting optimiser has to navigate during its run. This is shown in figure~\ref{fig:traces-vs-modes-EQ2}(a) with two straight line estimating the rate of decrease of the residual for $M=64$ and $M=8$. The case with more modes has a power law scaling equal to roughly $-2.6$ and the case with fewer modes has a power law of roughly $-3.4$. The larger value for $M=8$ indicates a faster rate of decrease of the residual. The case of $M=64$ resolvent modes achieved its final residual after roughly $10^5$ iterations, whereas the $M=8$ case shows the same residual after two orders of magnitude fewer of iterations. The figure clearly demonstrates the truncation of the resolvent modes, derived about the laminar base flow, associated with the smallest singular values improves convergence. This provides confirmation that the condition number of the Hessian at the minimum is improved, even though the resolvent modes are derived relative to a different base flow. A final note is that an immediate result of truncating the resolvent modes is to reduce the global residual of the initial field. This is a result of the noise introduced in the perturbation of S3 being truncated in the projection, reducing the smaller scale noise present in the starting flow. The initially smaller residual should not be the reason attributed to why the cases with fewer modes converge earlier, as the faster convergence is primarily associated with the faster convergence rate as discussed earlier.

In addition to the convergence rates, figure~\ref{fig:traces-vs-modes-EQ2}(b) shows the accuracy of the resulting solutions obtained from the projected optimisations plotted against the  number of resolvent modes used for the expansions in \eqref{eq:velocity-expansion} and \eqref{eq:residual-expansion}. The accuracy was computed as the norm of the difference between the field obtained from the projected optimisation, denoted with $\bm{u}_M$, and a base solution, denoted with $\bm{u}_\text{base}$, which was taken as the solution obtained using the full set of resolvent modes, i.e. $M=64$. As the number of modes used for the projection is increased, the error between the obtained solution and the base decreases, initially quite quickly when the number of modes used is small, but then reducing slower and saturating at around $\norm{\bm{u}_M-\bm{u}_\text{base}}^2_{\Omega_t}\approx10^{-7}$ at roughly $M=20$. The initial decrease in the error is a result of the extra modes being added having a relatively large contribution to the solution S3, with the observed exponential decrease in error typical of spectral methods. The saturation of the solution error as more modes are added is due to tolerances associated with the small changes in the minimum position and initial starting point for the optimisation for each value $M$.

\begin{figure}
    \centering
    \includegraphics{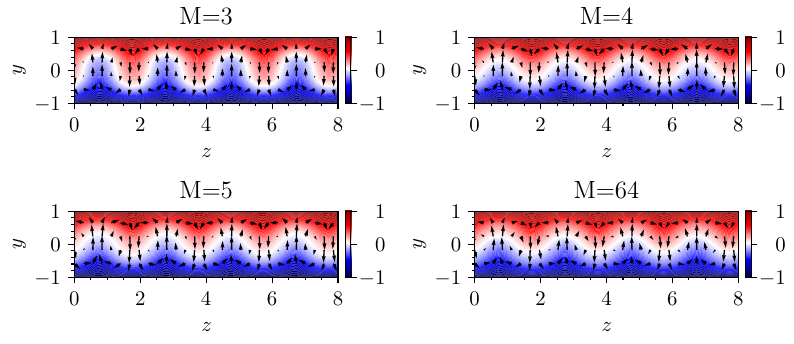}
    \caption{Snapshots of the solutions obtained from the projected optimisation of the perturbed S3 solution.  Contours represent the streamwise velocity and vectors represent the wall-normal and spanwise velocities.}
    \label{fig:EQ2-ROM-snapshots}
\end{figure}

Figure~\ref{fig:EQ2-ROM-snapshots} demonstrates that even a small number of resolvent modes can faithfully reconstruct the large-scale structure of a solution, despite potentially poor quantitative accuracy. It shows snapshots of the solutions obtained using $M=3$, $M=4$, $M=5$, and the base case $M=64$. Each solution displays the desired streamwise rolls with the correct spanwise wavelength. The solution for $M=3$ has noticeable qualitative differences from the base solution, consistent with its $\mathcal{O}\left(1\right)$ error in figure~\ref{fig:traces-vs-modes-EQ2}(b). However, the solutions for $M=4$ and $M=5$ are far more similar to the base solution visually, even though their quantitative error remains significant.

Taking these results together, we conclude that retaining a sufficiently small number of modes allows the projected optimisation to reconstruct the dominant structures of the desired solution, albeit at the cost of final accuracy, but with a significantly improved convergence rate. This presents a practical trade-off. If only the large-scale structures are sought, aggressively truncating the number of modes yields a solution far more rapidly. This could be used to initialise searches for ECS, using a truncated set of modes to quickly reduce the residual and capture the large-scale flow. The number of modes could then be increased to resolve smaller scales, or the output could be handed off to a Newton-GMRES-hookstep method for final convergence, akin to the approach in \citet{farazmand2016}.

\section{Conclusions}\label{sec:conclusions}
This work makes two primary contributions to the variational optimisation of exact coherent structures. First, we introduce a methodology for treating wall-bounded time periodic flows by projecting the optimisation onto a basis of divergence-free resolvent modes that inherently enforce no-slip boundary conditions and incompressibility. This approach, which extends the methods of \citet{farazmand2016} and \citet{schneider2022}, directly embeds the required constraints, alleviating a key theoretical difficulty. This is in contrast to the approach taken in \citet{schneider2023} using the influence matrix method. In doing so, it unifies the optimisation framework of these earlier works for finding ECSs with the resolvent-based modelling framework of \citet{mckeon2010}, providing a direct method to ``close the loop'' in resolvent analysis \citep{barthel2021} by solving for the self-consistent, finite-amplitude velocity field that sustains the chosen forcing modes. Second, we establish a formal link between the conditioning of the optimisation problem and resolvent analysis via the Hessian operator. We demonstrate that truncating the resolvent basis not only acts as an effective preconditioner for the optimiser, but also benefits from the resulting reduced-order model reducing the dimensionality of the problem while still leading to physically meaningful results.

The projected optimisation methodology is applied to 2-dimensional, 3-component rotating plane Couette flow, with all analysis performed with a rotation number of $\mathit{Ro}=0.5$. Both equilibria and periodic solutions are sought in this work. To ensure only minimal a priori knowledge about the flow is required the laminar profile is used to generate the required resolvent modes instead of the turbulent mean as is more standard for typical resolvent analysis applications. The implications of this choice and the effect it has on the modes has not been investigated here and warrants further study. The optimiser successfully identified equilibria that are not observed in DNS at $\mathit{Re}=50$ from various initial conditions. At $\mathit{Re}=400$, a stable periodic solution was obtained by initialising the optimiser with a perturbed DNS solution. In this case the residual decreased from $\mathcal{R}>10^{-3}$ to $\mathcal{R}<10^{-12}$ in roughly $40{,}000$ iterations, over half of which occurred in the first $5{,}000$. The slower convergence for the optimisation of the periodic solution is argued to be primarily a result of the increase in the  degrees of freedom of the system and the possible weak linear stability of the solution.  A notable omission is that of any time-periodic solutions obtained from optimisation that were not observed in DNS. Such results are omitted due to the very large cost of obtaining such solutions as the Reynolds number and computational domain size increase, specifically in terms of memory usage. To make these larger problems more feasible, a distributed implementation is required, which is discussed further below.

We also investigated the factors that affect the asymptotic convergence rate of the optimisation via the conditioning of the  discretised Hessian operator at a global minimum. It is shown that the closer a solution is to being marginally stable the larger the condition number and thus the slower the convergence, a behaviour which has been discussed in \citet{farazmand2016}. In addition, a direct link is established between the condition number of the Hessian and the resolvent expansion at a global minimum. It is shown that truncating the resolvent expansion by excluding the smallest singular values is equivalent to removing the fastest growing directions in the residual which optimally reduces the condition number of the Hessian operator. The improved convergence is demonstrated by optimising to an equilibrium solution using a reduced number of resolvent modes. Physically, this equivalence arises because the the Laplace operator dominates the higher-order behaviour of the Hessian eigenmodes and the resolvent modes regardless of the base flow around which the linearisations is performed. This operator is primarily responsible for the degradation in convergence rates, owing to its rapid dissipation of high-frequency components, particularly at higher Reynolds numbers where small-scale motions must be resolved. Removing the highest order resolvent modes is equivalent to removing these modes of growth from the optimisation. The resulting reduced-order solutions obtained from the optimisation on a truncated set of modes were shown to closely approximate the original solution derived from the full basis set. Specifically only a handful of modes were shown to be very close to the original solution. This means that even though the solution obtained using a very reduced set of modes does not accurately reconstruct the true solution in state-space, it could be used to initialise a more accurate search with superior convergence properties such as the Newton-GMRES-hookstep method.

Ultimately, the Galerkin projection not only enables the construction of solutions for general wall-bounded flows but also enhances the capability of variational optimisation. Coupled with its robustness to initial conditions, the improved convergence makes the method more practical both as a stand-alone solver and as a preconditioning step for root-finding approaches. The ability to truncate the modal basis further extends applicability to larger problems than are typically tractable with global solvers.

The projected variational optimiser presented in this work has two primary limitations. The most immediate is its high memory requirement for storing the high-dimensional state vector, which can become prohibitive for extending this method to fully three-dimensional turbulent flows. A promising path to mitigate this is leveraging parallel computing architectures, for which this framework is exceptionally well-suited since, unlike traditional direct numerical simulation of turbulence, the optimisation can be parallelised in the temporal dimension in addition to the spatial ones. A highly parallel implementation of the methodology would also permit an investigation into so-called ``quasi-trajectories'' from \citet{burton2025} for turbulent flows. These quasi-trajectories would constitute very low-order representations of the flow, similar in spirit to those explored by \citet{li2025}. However, whereas \citet{li2025} considered cases where the final objective was driven to very small values, sometimes near machine precision, the quasi-trajectories discussed by \citet{burton2025} would not aim to satisfy the low-order projection of the governing equations to such an extent. The second, more fundamental limitation is the efficacy of the resolvent basis being intrinsically linked to the chosen base flow. Determining a general principle for selecting the most effective base flow to yield useful, dynamically relevant modes requires extra investigation and represents an important direction for future work.

\begin{bmhead}[Supplementary movies.]
    Supplementary movies are available at
\end{bmhead}

\begin{bmhead}[Funding.]
    This work was supported using the Antony Wright Scholarship provided by the Department of Aeronautics and Astronautics at the University of Southampton.
\end{bmhead}

\begin{bmhead}[Competing interests.]
    The authors declare none.
\end{bmhead}

\begin{bmhead}[Author ORCIDs.\\]
    Thomas Burton \href{https://orcid.org/0000-0001-7998-2278}{https://orcid.org/0000-0001-7998-2278}; \\
    Sean Symon \href{https://orcid.org/0000-0001-9085-0778}{https://orcid.org/0000-0001-9085-0778}; \\
    Davide Lasagna \href{https://orcid.org/0000-0002-6501-6041}{https://orcid.org/0000-0002-6501-6041}; \\
\end{bmhead}

\appendix

\section{Residual Gradient Derivation}\label{app:residual-grad}

\subsection{Velocity Field Derivative}\label{app:residual-grad-wrt-velocity}
Before beginning the derivation, we shall define a modified optimisation problem that enforces the incompressibility constraint through a Lagrange multiplier
\begin{equation}
    \min_{\bm{u}\in\mathcal{P}_{T,\text{BC}},\,T}\quad\mathcal{R}\left[\bm{u},\,q\right]=\frac{1}{2}\norm{\bm{r}}^2_{\Omega_t}+\innprod{q}{\div{\bm{u}}}_{\Omega_t},
    \label{eq:global-residual-lagrange}
\end{equation}
where
\begin{equation}
    \mathcal{P}_{T,\text{BC}}=\left\{\bm{u}\in\chi\,\Big|\,\left.\bm{u}\right|_{t=0}=\left.\bm{u}\right|_{t=T},\,\left.\bm{u}\right|_{y=\pm1}=\pm\bm{e}_x,\,\bm{u}\text{ obeys periodic BCs}\right\},
\end{equation}
is the space of velocity fields that obey the boundary conditions, but are not necessarily incompressible. Also define the space of fields that obey the homogeneous boundary conditions
\begin{equation}
    \mathcal{P}_{T,\text{HBC}}=\left\{\bm{u}\in\chi\,\Big|\,\left.\bm{u}\right|_{t=0}=\left.\bm{u}\right|_{t=T},\,\left.\bm{u}\right|_{y=\pm1}=0,\,\bm{u}\text{ obeys periodic BCs}\right\},
\end{equation}
which will be useful for the later parts of this derivation. To derive the gradient first define a perturbation to the velocity field $\bm{u}\rightarrow\bm{u}+\epsilon\bm{v}$ where $0<\epsilon\in\mathbb{R}$ is small enough such that $\epsilon^2\approx0$. The perturbation is incompressible and obeys the homogeneous boundary conditions, i.e. $\bm{v}\in\mathcal{P}_{T,\text{HBC}}$ and $\div{\bm{v}}=0$. Calculus of variations provides the following identity
\begin{equation}
    \left[\dv{\epsilon}\mathcal{R}\left[\bm{u}+\epsilon\bm{v}\right]\right]_{\epsilon=0}=\innprod{\fdv{\mathcal{R}}{\bm{u}}}{\bm{v}}_{\Omega_t}.
    \label{eq:functional-derivative1}
\end{equation}
To derive the expression for $\fdv*{\mathcal{R}}{\bm{u}}$ it is required to propagate the perturbation $\bm{v}$ through all the variables that depend on $\bm{u}$. The local residual perturbation is
\begin{equation*}
    \bm{r}\left(\bm{u}+\epsilon\bm{v}\right)=\bm{r}+\bm{\delta r}=\pdv{\bm{u}}{t}+\epsilon\pdv{\bm{v}}{t}-\mathcal{N}\left(\bm{u}\right)-\delta\mathcal{N}-\grad{p}-\epsilon\grad{\delta p},
\end{equation*}
where
\begin{equation}
    \mathcal{N}\left(\bm{u}\right)=-\left(\bm{u}\cdot\bm{\nabla}\right)\bm{u}+\frac{1}{\mathit{Re}}\bm{\Delta}\bm{u}.
\end{equation}
Rearranging for $\bm{\delta r}$ gives
\begin{equation}
    \bm{\delta r}=\epsilon\left(\pdv{\bm{v}}{t}-\grad{\delta p}\right)-\delta\mathcal{N}.
    \label{eq:local-residual-perturbed}
\end{equation}
The perturbation term $\delta\mathcal{N}$ can be computed as
\begin{align*}
    \mathcal{N}\left(\bm{u}+\epsilon\bm{v}\right)&=\mathcal{N}+\delta\mathcal{N} \\
    =&-\left(\left(\bm{u}+\epsilon\bm{v}\right)\cdot\grad\right)\left(\bm{u}+\epsilon\bm{v}\right)+\frac{1}{\mathit{Re}}\bm{\Delta}\left(\bm{u}+\epsilon\bm{v}\right) \\
    =&-\left(\bm{u}\cdot\grad\right)\bm{u}+\frac{1}{\mathit{Re}}\bm{\Delta}\bm{u}-\epsilon\left(-\left(\bm{u}\cdot\grad\right)\bm{v}-\left(\bm{v}\cdot\grad\right)\bm{u}+\frac{1}{\mathit{Re}}\bm{\Delta}\bm{v}\right)+\mathcal{O}\left(\epsilon^2\right),
    \label{eq:ns-perturbed}
\end{align*}
which when rearranged for $\delta\mathcal{N}$, and neglecting the $\epsilon^2$ term, gives
\begin{equation}
    \delta\mathcal{N}=\epsilon\left(-\left(\bm{u}\cdot\grad\right)\bm{v}-\left(\bm{v}\cdot\grad\right)\bm{u}+\frac{1}{\mathit{Re}}\bm{\Delta}\bm{v}\right)+\mathcal{O}\left(\epsilon^2\right).
    \label{eq:ns-operator-perturbed}
\end{equation}
The perturbation in the global residual defined in \eqref{eq:global-residual-lagrange}, called the first variation in $\mathcal{R}$, is given by
\begin{align*}
    \mathcal{R}\left[\bm{u}+\epsilon\bm{v}\right]&=\frac{1}{2}\norm{\bm{r}+\bm{\delta r}}^2_{\Omega_t}+\innprod{q}{\div{\left(\bm{u}+\epsilon\bm{v}\right)}}_{\Omega_t} \\
    &=\frac{1}{2}\norm{r}^2_{\Omega_t}+\innprod{\bm{r}}{\bm{\delta r}}_{\Omega_t}+\frac{1}{2}\innprod{\bm{\delta r}}{\bm{\delta r}}_{\Omega_t}+\innprod{q}{\div{\bm{u}}}_{\Omega_t}+\epsilon\innprod{q}{\div{\bm{v}}}_{\Omega_t}. \\
\end{align*}
Substituting \eqref{eq:local-residual-perturbed} and \eqref{eq:ns-operator-perturbed} into the expression above, neglecting all the $\epsilon^2$ terms, gives
\begin{multline}
    \mathcal{R}\left[\bm{u}+\epsilon\bm{v}\right]=\frac{1}{2}\norm{\bm{r}}^2_{\Omega_t}+\innprod{q}{\div{\bm{u}}}_{\Omega_t}\\
    +\epsilon\Bigg(\innprod{\bm{r}}{\pdv{\bm{v}}{t}-\left(\bm{u}\cdot\grad\right)\bm{v}-\left(\bm{v}\cdot\grad\right)\bm{u}+\frac{1}{\mathit{Re}}\bm{\Delta}\bm{v}-\grad{\delta p}}_{\Omega_t}
    +\innprod{q}{\div{\bm{v}}}_{\Omega_t}\Bigg)+\mathcal{O}\left(\epsilon^2\right).
    \label{eq:global-residual-perturbed}
\end{multline}
Then, substituting \eqref{eq:global-residual-perturbed} into \eqref{eq:functional-derivative1} provides a relationship for the functional derivative
\begin{equation}
    \innprod{\fdv{\mathcal{R}}{\bm{u}}}{\bm{v}}_{\Omega_t}=\innprod{\bm{r}}{\pdv{\bm{v}}{t}-\left(\bm{u}\cdot\grad\right)\bm{v}-\left(\bm{v}\cdot\grad\right)\bm{u}+\frac{1}{\mathit{Re}}\bm{\Delta}\bm{v}-\grad{\delta p}}_{\Omega_t}+\innprod{q}{\div{\bm{v}}}_{\Omega_t}.
    \label{eq:functional-derivative2}
\end{equation}
To obtain a closed-form expression for $\fdv*{\mathcal{R}}{\bm{u}}$ it is necessary to derive the adjoint of all the operators on the right-hand side of \eqref{eq:functional-derivative2} such that it resembles the left-hand side. Starting with the time derivative
\begin{equation}
    \innprod{\bm{r}}{\pdv{\bm{v}}{t}}_{\Omega_t}=\innprod{-\pdv{\bm{r}}{t}}{\bm{v}}_{\Omega_t},
\end{equation}
where integration by parts has been used, with the boundary terms (at $t=0$ and $t=T$) cancel due to the flow being periodic in time. To simplify the derivation, we will make use of the product rule for divergence
\begin{equation}
    \div{(u\bm{v})}=u\div{\bm{v}}+\bm{v}\vdot\grad{u},
    \label{eq:product-rule}
\end{equation}
as well as the divergence theorem
\begin{equation}
    \int_\Omega\div{\bm{u}}\dd{V}=\int_{\partial\Omega}\bm{u}\vdot\vu{n}\dd{S},
    \label{eq:divergence-theorem}
\end{equation}
where $\vu{n}$ is an outwardly pointing normal vector on the surface $\partial\Omega$, and $\dd{V}$ and $\dd{S}$ are infinitesimal volume and surface elements, respectively. Note that for any field $\bm{u}\in\mathcal{P}_{T,\text{HBC}}$ the surface integral vanishes, i.e.
\begin{equation}
    \int_{\partial\Omega}\bm{u}\vdot\vu{n}\dd{S}=0,
\end{equation}
due to all the periodic boundary terms cancelling, and the homogeneous no-slip boundary terms being equal to zero. Thus, using \eqref{eq:product-rule} and \eqref{eq:divergence-theorem}, for the first of the convective terms we have
\begin{equation}
    \innprod{\bm{r}}{\left(\bm{u}\cdot\grad\right)\bm{v}}_{\Omega_t}=\int_0^T\left(\int_{\partial\Omega}\left(\bm{r}\cdot\bm{v}\right)\bm{u}\cdot\vu{n}\dd S-\int_\Omega\left(\left(\bm{u}\cdot\grad\right)\bm{r}\right)\cdot\bm{v}\dd{V}\right)\dd{t}.
\end{equation}
Using the fact that $\bm{v}\in\mathcal{P}_{T,\text{HBC}}$ and $\bm{u}\in\mathcal{P}_{T,\text{BC}}$, the surface integral vanish, leaving the following
\begin{equation}
    \innprod{\bm{r}}{\left(\bm{u}\cdot\grad\right)\bm{v}}_{\Omega_t}=\innprod{-\left(\bm{u}\cdot\grad\right)\bm{r}}{\bm{v}}_{\Omega_t}.
\end{equation}
The second convective term is slightly simpler due to $\bm{v}$ not appearing in any derivatives
\begin{equation}
    \innprod{\bm{r}}{\left(\bm{v}\cdot\grad\right)\bm{u}}_{\Omega_t}=\innprod{\bm{r}}{\left(\grad{\bm{u}}\right)\bm{v}}_{\Omega_t}=\innprod{\left(\grad{\bm{u}}\right)^\top\bm{r}}{\bm{v}}_{\Omega_t},
\end{equation}
where $\grad{\bm{u}}$ is the gradient of the vector field $\bm{u}$. In Cartesian coordinates this looks like the gradient of the scalar components on $\bm{u}$ stacked on top of each other to form a matrix. Next, the diffusion term
\begin{equation}
    \innprod{\bm{r}}{\bm{\Delta}\bm{v}}_{\Omega_t}=\int_0^T\left(\int_{\partial\Omega}\left(\bm{r}\cdot\grad{\bm{v}}-\grad{\bm{r}}\cdot\bm{v}\right)\cdot\vu{n}\dd S+\int_\Omega\bm{\Delta}{\bm{r}}\cdot\bm{v}\dd V\right)\dd t,
\end{equation}
where the product rule and divergence theorem have been used twice to effectively integrate by parts the second-order derivatives in the Laplacian. The second of the boundary integrals vanish due to $\bm{v}\in\mathcal{P}_{T,\text{HMB}}$, however, the other boundary integral does not by default since we have not imposed any extra restrictions on the gradients of $\bm{v}$. To resolve this, it is necessary to impose $\bm{r}\in\mathcal{P}_{T,\text{HBC}}$, enforcing that $\bm{r}$ vanishes at the walls, which means the first boundary integral also vanishes, giving
\begin{equation}
    \innprod{\bm{r}}{\bm{\Delta}\bm{v}}_{\Omega_t}=\innprod{\bm{\Delta}\bm{r}}{\bm{v}}_{\Omega_t}.
\end{equation}
Moving to the pressure gradient we have
\begin{equation}
    \innprod{\bm{r}}{\grad{\delta p}}_{\Omega_t}=\int_0^T\left(\int_{\partial\Omega}\delta p\left(\bm{r}\vdot\vu{n}\right)\dd S-\int_\Omega\left(\div{\bm{r}}\right)\delta p\dd{V}\right)\dd{t}.
\end{equation}
Since we know that $\bm{r}\in\mathcal{P}_{T,\text{HBC}}$ as prescribed by the adjoint of the diffusion term, the boundary term here must vanish, leaving
\begin{equation}
    \innprod{\bm{r}}{\grad{\delta p}}_{\Omega_t}=\innprod{-\div{\bm{r}}}{\delta p}_{\Omega_t}.
    \label{eq:pressure-gradient}
\end{equation}
The divergence of the residual is actually the  pressure-Poisson equation, and since the pressure field is assumed to satisfy this equation we also know that $\div{\bm{r}}=0$, and thus \eqref{eq:pressure-gradient} is also zero. This just leaves the divergence term for the Lagrange multiplier, enforcing the divergence-free evolution of the velocity field $\bm{u}$ under the optimisation problem given in \eqref{eq:global-residual-lagrange}. Due to the symmetry of the  inner product (in the case of real fields), it is easy to see that the adjoint of the divergence operator is the negative gradient operator. Applying this result gives
\begin{equation}
    \innprod{q}{\div{\bm{v}}}_{\Omega_t}=\int_0^T\int_\Omega-\grad{q}\cdot\bm{v}\dd V\dd t=\innprod{-\grad{q}}{\bm{v}}_{\Omega_t}.
\end{equation}
Combining all of these results and substituting them into the right-hand side of \eqref{eq:functional-derivative2} results in
\begin{equation}
    \innprod{\fdv{\mathcal{R}}{\bm{u}}}{\bm{v}}_{\Omega_t}=\innprod{-\pdv{\bm{r}}{t}-\left(\bm{u}\cdot\grad\right)\bm{r}+\left(\grad{\bm{u}}\right)^\top\bm{r}}-\frac{1}{\mathit{Re}}\bm{\Delta}\bm{r}-\grad{q}{\bm{v}}_{\Omega_t},
\end{equation}
which provides the desired closed-form expression for the functional derivative of $\mathcal{R}$ with respect to $\bm{u}$
\begin{equation}
    \fdv{\mathcal{R}}{\bm{u}}=-\pdv{\bm{r}}{t}-\left(\bm{u}\cdot\grad\right)\bm{r}+\left(\grad{\bm{u}}\right)^\top\bm{r}-\frac{1}{\mathit{Re}}\bm{\Delta}\bm{r}-\grad{q},
    \label{eq:residual-gradient}
\end{equation}
with the additional constraints $\bm{r}\in\mathcal{P}_{T,\text{HBC}}$ and $\div{\bm{r}}=0$. The boundary constraints on $\bm{r}$ at the boundaries is an instance of a natural boundary conditions, and are necessary to enforce. If they are not strictly enforced, then the residual gradient given in \eqref{eq:residual-gradient} is not guaranteed to be a descent direction. A similar interpretation applies to $\div{\bm{r}}=0$ except it applies throughout the whole domain rather than just the boundaries.

\subsection{Fundamental Frequency Derivative}\label{app:residual-grad-wrt-frequency}
The derivative of $\mathcal{R}$ with respect to the fundamental frequency $\omega$ is simpler to derive since $\mathcal{R}$ is an ordinary function of $\omega$ rather than a functional such as with $\bm{u}$. First, define a new scaled time $t^\prime=\omega t$ which then leads to a the following relationship between the time derivatives $\pdv*{t}=\omega\pdv*{t^\prime}$. Using this scaled time in the definition of the local residual in \eqref{eq:residual}, and then substituting the resulting expression into the definition of the global residual gives
\begin{multline}
    \mathcal{R}=\frac{\omega^2}{2}\norm{\pdv{\bm{u}}{{t^\prime}}}^2_{\Omega_t}-\omega\innprod{\pdv{\bm{u}}{t^\prime}}{-\left(\bm{u}\vdot\grad{\bm{u}}\right)+\frac{1}{\mathit{Re}}\bm{v}-\grad{p}}_{\Omega_t}\\+\frac{1}{2}\norm{-\left(\bm{u}\vdot\grad{\bm{u}}\right)+\frac{1}{\mathit{Re}}\bm{v}-\grad{p}}^2_{\Omega_t}.
    \label{eq:global-residual-wrt-frequency}
\end{multline}
Fixing the particular velocity field $\bm{u}$ under consideration makes this a quadratic function, with a single global minimum at some $\omega^*$. This global minimum could be derived rather simply, however using it directly during the optimisation was found to lead to very erratic behaviour, especially when far away from a solution. Taking the derivative of \eqref{eq:global-residual-wrt-frequency} with respect to $\omega$ gives
\begin{equation}
    \pdv{\mathcal{R}}{\omega}=\omega\norm{\pdv{\bm{u}}{t^\prime}}^2_{\Omega_t}-\innprod{\pdv{\bm{u}}{t^\prime}}{-\left(\bm{u}\vdot\grad{\bm{u}}\right)+\frac{1}{\mathit{Re}}\bm{v}-\grad{p}}_{\Omega_t}.
\end{equation}
With a slight rearrangement the following simplified expression is obtained
\begin{equation}
    \pdv{\mathcal{R}}{\omega}=\frac{1}{\omega}\innprod{\pdv{\bm{u}}{t}}{\bm{r}}_{\Omega_t}.
\end{equation}
A similar expression can be obtained for the derivative with respect to the solution period $T=2\pi/\omega$ using the chain rule
\begin{equation}
    \pdv{\mathcal{R}}{T}=-\frac{1}{T}\innprod{\pdv{\bm{u}}{t}}{\bm{r}}_{\Omega_t}.
\end{equation}

\section{DAE Form of Variational Optimisation}\label{app:dae}
In the literature \citep{farazmand2016,schneider2022,schneider2023} the variational optimisation is derived using an adjoint variational dynamics formulation. Using the same notation, where $\tau$ represents the fictitious time introduced to parametrise the evolution of the variational dynamics, we can say that
\begin{equation}
    \pdv{\bm{u}}{\tau}:=-\fdv{\mathcal{R}}{\bm{u}},\quad\dv{\omega}{\tau}:=-\pdv{\mathcal{R}}{\omega},
\end{equation}
Using this definition the gradient-based optimisation problem can be restated in the form of a Differential-Algebraic Equation (DAE) for the evolution of the spatio-temporal flow field
\begin{subequations}
    \begin{align}
        \pdv{\bm{u}}{\tau}&=\pdv{\bm{r}}{t}+\left(\bm{u}\vdot\grad\right)\bm{r}-\left(\grad{\bm{u}}\right)^\top\bm{r}+\frac{1}{\mathit{Re}}\bm{\Delta}\bm{r}-\grad{q}, \label{eq:dae1}\\
        \pdv{\omega}{\tau}&=-\frac{1}{\omega}\innprod{\pdv{\bm{u}}{t}}{\bm{r}}_{\Omega_t}, \label{eq:dae-omega}\\
        0&=\pdv{\bm{u}}{t}+\grad{p}+\left(\bm{u}\cdot\bm{\nabla}\right)\bm{u}-\frac{1}{\mathit{Re}}\bm{\Delta}\bm{u}-\bm{r}, \label{eq:dae2}\\
        0&=\div{\bm{u}}, \label{eq:dae3}\\
        0&=\div{\bm{r}}, \label{eq:dae4}\\
        0&=\left.\bm{u}\right|_{y=\pm1}\mp\bm{e}_x, \label{eq:dae-ns1}\\
        0&=\left.\bm{r}\right|_{y=\pm1}, \label{eq:dae-ns2}\\
    \end{align}
    \label{eq:dae}%
\end{subequations}
In the work of \citet{schneider2022,schneider2023}, equation \eqref{eq:dae} is solved using a simple first-order Euler scheme to explicitly step forward the solution by $\Delta\tau$. The use of the Euler method was justified as the accuracy of the intermediate result does not matter, just the accuracy of the solution at the end once $\mathcal{R}$ becomes sufficiently small. This allowed for larger step sizes in $\tau$. \citet{farazmand2016} uses a Runge-Kutta time stepping method in $\tau$ to improve accuracy, but this still suffers from the same slow convergence expected from gradient descent methods. However, as shown here, the problem can be viewed simply as an optimisation problem, where the goal is to minimise $\mathcal{R}$ as much as possible regardless of the route taken in the loop space to do so. This means the intermediate accuracy of any time-stepping scheme in $\Delta\tau$ is not productive. Instead, more broad optimisation algorithms can be used, that trade the exact evolution prescribed by \eqref{eq:dae1} and \eqref{eq:dae-omega} for convergence to smaller residuals in far fewer iterations. This is displayed in section~\ref{sec:equilibria} where gradient descent (mathematically equivalent to the Euler method in $\tau$) is compared with common quasi-Newton optimisation algorithms.

\section{Resolvent for 2D3C RPCF}\label{app:resolvent}
In this work we use resolvent analysis to generate a divergence-free and no-slip orthonormal basis set for the velocity and local residual fields. The exact form of the operators defined in section~\ref{sec:resolvent} for 2D3C RPCF is given here. First, the (nonlinear) Navier-Stokes operator is defined as
\begin{equation}
    \mathcal{N}\left(\bm{u}_b\right)=-\left(\bm{u}_b\vdot\grad\right)\bm{u}_b+\dfrac{1}{\mathit{Re}}\bm{\Delta}\bm{u}_b-\mathit{Ro}\left(\bm{e}_z\cross\bm{u}_b\right),
\end{equation}
where the base pressure field $p_b$ is not included since it is projected away by the Leray projector $\mathbb{P}$ in \eqref{eq:ns-fluc}. The linearised Navier-Stokes operator evaluated at the base flow $\bm{u}_b$, $\mathcal{L}_{\bm{u}_b}$, is given by
\begin{equation}
    \mathcal{L}_{\bm{u}_b}\bm{u}^\prime=-\left(\bm{u}_b\vdot\grad\right)\bm{u}^\prime-\left(\grad{\bm{u}_b}\right)\bm{u}^\prime+\frac{1}{\mathit{Re}}\bm{\Delta}\bm{u}^\prime-\mathit{Ro}\,\bm{e}_z\cross\bm{u}^\prime.
\end{equation}
where again the pressure is not included due to the Leray projector in \eqref{eq:ns-fluc}. Assuming  $\bm{u}_b(y)=u_b(y)\bm{e}_x$, and using the definition of the resolvent in section~\ref{sec:resolvent}, the final expression for the operator for 2D3C RPCF is expressed as
\begin{equation}
    \bm{R}_{\bm{k}}=\left(\begin{array}{ccc}
        ik_t\omega-\dfrac{1}{\mathit{Re}}\bm{\Delta}_{\bm{k}} & \dfrac{\partial u_b}{\partial y}-\mathit{Ro} & 0 \\
        \mathit{Ro} & ik_t\omega-\dfrac{1}{\mathit{Re}}\bm{\Delta}_{\bm{k}} & 0 \\
        0 & 0 & ik_t\omega-\dfrac{1}{\mathit{Re}}\bm{\Delta}_{\bm{k}}
    \end{array}\right)^{-1}\mathbb{P}.
\end{equation}
where $\bm{\Delta}_{\bm{k}}=\partial^2/\partial y^2-(\beta k_z)^2$.

\section{DNS Numerical Methods}\label{app:dns-solver}
The velocity vector field is conveniently decomposed into the laminar base state, and the velocity fluctuation denoted in this appendix as  $\bm{u}(y,\,z,\,t)=\left(u,\,v,\,w\right)$. A formulation using the streamwise component of the vorticity perturbation $\displaystyle \omega_x = \partial w / \partial y - \partial v / \partial z$, and the streamwise (fluctuation) velocity $u$, is used to numerically solve the problem \eqref{eq:ns}. This formulation eliminates complications arising from the pressure term. Because the domain is two-dimensional the stream function $\psi$, related to the transverse and wall-normal (fluctuation) velocity components by the relations $w = -\partial \psi/\partial y$ and $v = {\partial \psi}/{\partial z}$, and to the vorticity by the Poisson-type equation $\bm{\Delta}\psi = - \omega$, is introduced. Introduction of the streamfunction ensures that the continuity equation is satisfied. With these definitions, the coupled system of nonlinear Partial Differential Equations (PDEs)
\begin{subequations}
    \begin{align}
        \frac{\partial u}{\partial t} =& -\frac{\partial \psi}{\partial z}\frac{\partial u}{\partial y}+\frac{\partial \psi}{\partial y}\frac{\partial u}{\partial z}  + \frac{\partial \psi}{\partial z}(Ro-1) +\frac{1}{Re}\bm{\Delta}u, \\
        \frac{\partial \omega }{\partial t} =& - \frac{\partial \psi }{\partial z}\frac{\partial \omega }{\partial y} +\frac{\partial \psi}{\partial y}\frac{\partial \omega}{\partial z} -\frac{\partial u}{\partial z}Ro +\frac{1}{Re}\bm{\Delta}\omega,
    \end{align}
\label{eq:vort-form}%
\end{subequations}
governs the evolution of the streamwise (fluctuation) velocity and vorticity. A spectral spatial discretisation of \eqref{eq:vort-form} is adopted along the axial direction by introducing the truncated Fourier expansion
\begin{equation}\label{eq:fourier}
  \displaystyle f(z, y, t) = \sum_{k=-N/2}^{N/2-1} f_k(y, t) \exp( i k_z \beta z)
\end{equation}
for $u,\,\omega$ and $\psi$, with $\beta= 2\pi/L_z$, and with $k_z$ being the spanwise wavenumber. Projection of the governing equations \eqref{eq:vort-form} onto \eqref{eq:fourier} leads to a system of sets of three one-dimensional coupled partial differential equations in the wall-normal direction $y$, for each wave number $k_z$. This system of PDEs is then discretised in time using a fairly standard splitting approach. The viscous term is treated implicitly using a Crank-Nicholson method, while the nonlinear term and the term arising from the rotation and the mean flow shear are integrated explicitly via a second-order accurate Adams-Bashforth method. The nonlinear term is treated pseudo-spectrally, see \citet{canuto1988}, whereas de-aliasing was not found necessary, given the low resolution requirements. Temporal discretisation results in a system of ordinary differential equations in the space variable $y$, having the form of Helmoltz problems for each wave number $k_z$ and for the three flow variables. These are solved at each time step using a second-order centred finite difference scheme, leading to sparse banded system solved directly by a Gaussian elimination method with pivoting.

\section{Hessian Operator Derivation}\label{app:hessian-derivation}
To derive the Hessian operator, first take the definition of the local residual in \eqref{eq:residual} and project it onto the divergence-free subspace using the Leray projector from section~\ref{sec:resolvent}
\begin{equation}
    \bm{r}=\pdv{\bm{u}}{t}-\mathbb{P}\mathcal{N}\left(\bm{u}\right),
    \label{eq:residual-leray}
\end{equation}
where the local residual itself is unaffected by the projection since it is divergence-free by construction. Consider a solution to the optimisation problem denoted as $\bm{u}^*$. A perturbation to the solution $\bm{u}^*$ by some arbitrary velocity field $\bm{V}$ that is incompressible and obeys the appropriate boundary conditions is applied. The global residual $\mathcal{R}$ can then be expanded at the minimum using a Taylor series
\begin{equation}
    \mathcal{R}\left[\bm{u}^*+\epsilon\bm{v}\right]=\mathcal{R}\left[\bm{u}^*\right]+\epsilon\left\langle\frac{\delta\mathcal{R}}{\delta\bm{u}^*},\,\bm{v}\right\rangle_{\Omega_t}+\frac{\epsilon^2}{2}\left\langle\bm{v},\,\bm{H}\bm{v}\right\rangle_{\Omega_t}+\mathcal{O}\left(\epsilon^3\right)=\frac{\epsilon^2}{2}\left\langle\bm{v},\,\bm{H}\bm{v}\right\rangle_{\Omega_t}+\mathcal{O}\left(\epsilon^3\right),
\end{equation}
where the fact that $\bm{u}^*$ is a minimum means that $\mathcal{R}[\bm{u}^*]=\lVert\delta\mathcal{R}/\delta\bm{u}^*\rVert=0$. We then have a local quadratic model for the variation of $\mathcal{R}$ that is governed by the Hessian operator $\bm{H}$. The second-order variation of the global residual is given by $\mathcal{R}[\bm{u}^*+\epsilon\bm{v}]=\lVert\delta\bm{r}\rVert^2_{\Omega_t}/2$. Linearising the local residual in \eqref{eq:residual-leray} and substituting this into the previous expressions gives
\begin{equation}
    \left\langle\bm{v},\,\bm{H}\bm{v}\right\rangle_{\Omega_t}=\left\lVert\frac{\partial\bm{v}}{\partial t}-\mathbb{P}\mathcal{L}_{\bm{u}^*}\bm{v}\right\rVert^2_{\Omega_t}.
\end{equation}

\bibliographystyle{jfm}
\bibliography{export}



\end{document}